\documentclass[10pt,oneside,reqno]{amsart}

\usepackage{amstext,amsmath,amsfonts,amscd,amsthm,graphicx}

\theoremstyle{plain}
\newtheorem{theorem}{Theorem}
\newtheorem{lemma}[theorem]{Lemma}

\newtheorem{corollary}[theorem]{Corollary}

\newtheorem{definition}{Definition}

\newtheorem{example}{Example}

\theoremstyle{remark}
\newtheorem{remark}{Remark}
\newtheorem{note}{Note}
\newtheorem{joke}{Joke}

\newcommand{\dirac}{\not\!\!{D}}
\begin{document}
\title{A Spectral Quadruple for de Sitter Space}
\author{Tom\'{a}\v{s} Kopf}
\address{Matematical Institute of the Silesian University at Opava,
Bezru\v{c}ovo n\'{a}m\v{e}st\'{i} 13, 746 01 Opava, Czech Republic}
\curraddr{ThEP, Institut f\"{u}r Physik, Johannes
Gutenberg-Universit\"{a}t,
55099 Mainz, Germany}
\email{kopf@ThEP.Physik.Uni-Mainz.DE}
\thanks{First author supported by the Alexander von Humboldt Foundation}

\author{Mario Paschke}
\address{ThEP, Institut f\"{u}r Physik, Johannes
Gutenberg-Universit\"{a}t,
55099 Mainz, Germany}
\email{paschke@ThEP.Physik.Uni-Mainz.DE}


\date{November 21, 2000}



\begin{abstract}
A set of data supposed to give possible
axioms for spacetimes with a sufficient number of isometries in spectral
geometry is given. These data are shown to be sufficient to
obtain 1+1 dimensional de Sitter spacetime. The data rely at the moment somewhat on the
guidance given by a required symmetry, in part to allow
explicit calculations in a specific model.
The framework applies also to the noncommutative case. Finite spectral triples are discussed as an example.
\end{abstract}

\maketitle

\section{Introduction}

Spacetime is the fairy tale of a classical manifold. It is irreconcilable with
quantum effects in gravity and most likely, in a strict sense, it does not
exist. But to dismiss a mythical being that has inspired generations just
because it does not really exist is foolish. Rather it should be understood
together with the story-tellers through whom and in whom the being exists.

The story-tellers of spacetime are the physical fields. It is the interaction
with fields that lets one believe that there was a spacetime in which particles
of a field have propagated.

Here, the fairy tale of classical spacetime will be
uncritically retold again: no quantum effects will be considered and the result
will be an ordinary manifold of Lorentzian signature. And yet it is not at all
the same old story:  the story-tellers, the fields are put into the place
they deserve - at the centre of the matter.

The right framework to allow the fields to play their distinguished role is
noncommutative geometry, in particular the notion of a spectral
triple\cite{Connes94,Connes95,Connes96a}. A
spectral triple $(A, \mathcal{H}, D, J,\gamma)$ consists of a
pre-${C}^{\ast}$-algebra $A$ represented (faithfully) on a Hilbert space
$\mathcal{H}$, an unbounded selfadjoint operator $D$ on $\mathcal{H}$, an
antiunitary operator $J$ and of a grading operator $\gamma$ on
$\mathcal{H}$. These structures satisfy a set of seven conditions given in
\cite{Connes96a}:

\begin{enumerate}

\item{\bf Classical dimension.} The inverse ${D}^{-1}$ of $D$ is an
infinitesimal of order $\frac {1} {n}$.

\item{\bf First order condition.} For any $a$ in the algebra $A$ and $b$ in the
opposite algebra (represented with the help of $J$),
\begin{align}
[[D,a],{b}^{op}]&=0
\end{align}

\item{\bf Regularity.} The elements $a$ of the algebra $A$ as well as their
commutators $[D,a]$ with $D$ are smooth vectors of the derivation $[\mid D\mid ,\bullet
]$

\item{\bf Orientability.}
There exists the image $\gamma$ of a Hochschild cycle in degree n such that
\begin{align}
\intertext{for $n$ even}
\gamma &= {\gamma}^{\ast} & {\gamma}^{2} &=\mathbf{1} & \gamma D&=- D\gamma\\
\intertext{and for $n$ odd}
&&\gamma &=\mathbf{1}&&
\end{align}

\item{\bf Finiteness.} The subspace of smooth Hilbert space vectors
$\mathcal{H}_{\infty}$ is a finite projective module over $A$ with a local
hermitean structure.

\item{\bf Poincar\'{e} duality.} The intersection form ${K}_{\bullet}\times
{K}_{\bullet}\rightarrow\mathbb{Z}$ on the $K$-theory ${K}_{\bullet}$ is
invertible.

\item{\bf Reality.} $J$ satisfies
\begin{align}
{J}^{2}&={(-1)}^{\frac {(n-1)n(n+1)(n+2)} {8}} & JD&={(-1)}^{\frac {n(n+1)(n+2)} {2}}DJ\\
\intertext{and for $n$ even}
J\gamma &= {(-1)}^{\frac {n} {2}}\gamma J&&
\end{align}
\end{enumerate}

If the algebra $A$ is commutative then in a rather deep sense the
spectral triple is more or less the same as a spin manifold $M$
with positive definite metric \cite{Connes96a} that can be
recovered (for a proof see \cite{Rennie},
\cite{Gracia-Bondia-Varilly-Figueroa}) in the following way: the
algebra $A$ is the algebra of functions on the spin manifold,
$\mathcal{H}$ is the Hilbert space of square-integrable sections
of the spinor bundle over $M$, $D$ is the Dirac operator, $C$ is
the charge conjugation and $\gamma$ is the volume element. All
spin manifolds with positive definite metric can be obtained in
this way. A generalization covering the orientable Riemannian
case without a spin structure was recently given by S. Lord
\cite{Lord}.

The possibility of choosing a noncommutative algebra $A$ even allows to go beyond ordinary manifolds.

However, this description is not directly applicable to spin manifolds with
Lo\-rentz\-ian signature. One problem seems to be that in this case the canonical
inner product on spinor fields is not positive definite.

A way to deal with this is to foliate spacetime with spacelike leaves and
to describe the Euclidean geometry of the leaves as above. Under the assumption
that the spinor fields satisfy the Dirac equation, spinor fields on different
leaves belonging to the same solution can be identified \cite{Hawkins97}. Then
the algebras
$A(t)$
for different leaves are represented on the same Hilbert space and
their causal relationships may be tested by examination of their commutators.
This can be exploited to
obtain considerable information on the geometry and render further structures
typical for Hamiltonian approaches, e. g., the lapse and the shift superfluous
\cite{Kopf98}.

The Hilbert space $\mathcal{H}$ has in this picture a clear physical
interpretation as the
phase space of a spinor field. An interpretation of the algebras $A(t)$ was
suggested in
\cite{Kopf1999} and will be scrutinized in future work.

However, previous results did not show how to obtain meaningful spectral
information nor what
axioms to start with in order to get interesting spacetime geometries. This task
is
carried out here in a particular example leading to $1+1$-dimensional de Sitter spacetime.
Such an
undertaking is not easy even in the Riemannian case \cite{Paschke-thesis} where
a set of conditions as
reviewed above is available. Therefore, the realization that
postulated symmetries allow to completely carry out explicit calculations is an
important
ingredient of this work. This should however not obscure the fact that the
requirements of the
here discussed spectral quadruple {\em do suggest} what choices are to be made for a
general set of
axioms, in absence of any symmetries. At the same time this work provides a
model against
which any set of axioms may be tested.

This paper is structured as follows: Practical calculations to extract the
spacetime geometry of a globally hyperbolic spin manifold out of commutators
\cite{Kopf-Paschke2000} and the generators of evolution (Hamiltonians) are
reviewed in Section \ref{globalspacetime}. These motivate  the definition of a spectral
quadruple in Section \ref{spectralspacetime}. This definition is to provide definiteness rather than definitiveness and is
to be understood as a working hypothesis. In Section \ref{SL2Rreptheory},
the representation theory of the relevant symmetry group,
${SL}_{2}(\mathbb{R})$ is reviewed. The spacetime of the postulated  spectral
quadruple is calculated in Section \ref{deSitterrecover}. A discrete spectral
quadruple is discussed as a further example in Section
\ref{discretequadruples}. The Conclusion contains a discussion of the obtained
results and of possibilities for further investigations.  A collection of
calculations and formulae for 1+1 de Sitter spacetime, that might be useful for other purposes also, is contained in the
Appendix.


\section{The reconstruction of spacetime from commutators}\label{globalspacetime}


In this section, a globally hyperbolic $n$-dimensional spacetime $M$ with a given foliation ${\Sigma}_{t}$,
$t\in\mathbb{R}$ by spacelike Cauchy hypersurfaces are preassumed. This is sufficient to define a family ${A}_{t}$ of
commutative algebras, the algebras of smooth functions on the hypersurfaces ${\Sigma}_{t}$,
\begin{align}
{A}_{t} &= {C}^{\infty}({\Sigma}_{t}),
\end{align}
 and to consider their action on the classical phase space (the space of solutions) $\mathcal{H}$ of a Dirac spinor field
 satisfying the Dirac equation,

 \begin{align}
 (\dirac - m) \psi &= 0 &\text{for $\psi\in\mathcal{H}$.}\label{Diracequation}
 \end{align}
This action is just the pointwise multiplication of the initial data of the spinor field by functions on the corresponding
Cauchy space slice:

\begin{align}
({a}_{t}\psi)(x)&= {a}_{t}(x)\psi (x)&\text{for ${a}_{t}\in{A}_{t}$, $\psi\in\mathcal{H}$ and $x\in {\Sigma}_{t}$.}
\end{align}
It is well defined, since a solution is uniquely given by its initial values on any Cauchy hypersurface.

While the algebras ${A}_{t}$ are themselves commutative, they do not commute with each other. We proceed now, following
\cite{Kopf-Paschke2000}, to express this noncommutativity in terms of commutators and then show that the knowledge of the
calculated commutators is sufficient to recover the preassumed spacetime manifold including its metric structure and spin
structure.

The commutators of interest will be those of functions taken at different times and of functions with the (generically
time-dependent) Hamiltonian $H$ of the family ${A}_{t}$ of algebras. In the calculations, we will assume to be given the
standard data of the Hamiltonian description of gravity, the ADM (Arnowitt-Deser-Misner) formalism
\cite{Arnowitt-Deser-Misner}. A time flow
identifying the Cauchy surfaces in the family and the metric tensor ${g}_{ij}$ on the hypersurfaces describe the leaves.
Also needed is a lapse function $N$ describing the infinitesimal distance between
hypersurfaces and the shift function ${N}^{i}$ expressing, how much the identification of neighboring hypersurfaces (the
time vector $\frac {\partial} {\partial t}$) is shifted from the normal direction ${e}_{\bot}$ along the spacial
directions ${e}_{i}$:

\begin{align}
\frac {{\partial}^{\alpha}} {\partial t} &= N {e}_{\bot} + {N}^{i} {e}_{i}
\end{align}

\begin{figure}[h]
\begin{center}
\includegraphics[width=0.5\textwidth]%
{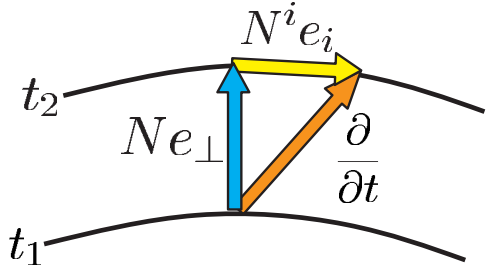}
\end{center}
\caption[]{{\bf The $n+1$ splitting of spacetime in the ADM
formalism.} The time vector $\frac {\partial} {\partial t}$
giving the evolution from a Cauchy hypersurface at time ${t}_{1}$
to the infinitesimally later Cauchy hypersurface at time
${t}_{2}$ can be decomposed into its part ${N}^{i} {e}_{i}$
parallel to the hypersurfaces and into its orthogonal part $N
{e}_{\bot}$. Here ${e}_{i}$, ${e}_{\bot}$ are unit vectors
adapted to the hypersurfaces and ${N}^{i}$, $N$ are the shift and
the lapse. } \label{adm}
\end{figure}

In addition, the Dirac generators ${\not \! e}_{\bot}$, ${\not \! e}_{i}$ corresponding to the covectors
${e}_{\bot}$, ${e}_{i}$ and the canonical spin connection $\omega$ will appear in the calculations. (Here, the common slash
notation is used: The natural insertion ${\gamma}^{\mu}$ of covectors ${a}_{\mu}$ into the Clifford algebra generated by
covectors is just indicated by a slash, ${\not\! a}:={\gamma}^{\mu}{a}_{\mu}$.)

With this, the Hamiltonian of the spinor field can be written as

\begin{align}\label{spinorHamiltonian}
iH &= -N\left( {\omega}_{\bot}^{S} + {\not \! e}_{\bot}{\not \! e}_{i}{g}^{ij}{D}_{j}
- m{\not \! e}_{\bot} \right) + {N}^{i} {\partial}_{i}.
\end{align}

It is obtained by $n+1$ splitting of the Dirac equation
(\ref{Diracequation})
\begin{align}
\left( -{\gamma}_{\bot}\left( \frac {1} {N} {D}_{t}-\frac {N^i}
{N} {D}_{i} \right) + {\gamma}_{i}{g}^{ij}{D}_{j}-m\right)\psi&=0,
\end{align}
multiplying from the right by ${\not \! e}_{\bot}$ and then
extracting the time derivative term. The form
(\ref{spinorHamiltonian}) is then obtained using the splitting of
the spin connection form in the time direction:
\begin{align}
{\omega}^{S}_{t}&=N{\omega}^{S}_{\bot}+{N}^{i}{\omega}^{S}_{i}.
\end{align}

The commutator of the Hamiltonian with a function $f=f({t}_{0})$ belonging to a particular algebra ${A}_{{t}_0}$ is easily computed:

\begin{align}
[ f, iH] &= N{\not \! e}_{\bot}{\not \! e}_{i}{g}^{ij}({\partial}_{j}f)\label{commutatorwithH}
-{N}^{i}{\partial}_{i}f
\end{align}

The commutator of two functions $g\in {\Sigma}_{{t}_{0}}$, $f\in {\Sigma}_{{t}_{1}}$ is a more subtle issue. A not so
short calculation, taking into account the possible time dependence of the Hamiltonian, gives the following expansion in
time ${t}_{1}-{t}_{0}$:
\begin{align}
[f({t}_{1}),g({t}_{0})] =& \,\,\, {({t}_{1}-{t}_{0})}^{0}0\label{zerothorder}\\
                        & +{({t}_{1}-{t}_{0})}^{1}0\label{firstorder}\\
                        & -{({t}_{1}-{t}_{0})}^{2}({N}^{2} {g}^{ij}{g}^{kl}({\partial}_{i}f)[{\not\! e}_{j},{\not\!
                        e}_{k}]{\partial}_{l}g)\label{secondorder}\\
                        & +{({t}_{1}-{t}_{0})}^{3}4{\not\! e}_{\bot}m{N}^{3}{g}^{ij}
                        ({\partial}_{i}f)({\partial}_{j}g)+
                        \begin{pmatrix} \text{terms vanishing}\\ \text{in $1+1$ dimensions}\end{pmatrix}\label{thirdorder}\\
                        & + O\left({({t}_{1}-{t}_{0})}^{4}\right)
\end{align}
\begin{note}
All functions on the right hand side are taken at the time ${t}_{0}$. It is understood that $f({t}_{0})\in {A}_{{t}_{0}}$
is the element corresponding to $f({t}_{1})\in {A}_{{t}_{1}}$ under the identification of  ${A}_{{t}_{0}}$,
${A}_{{t}_{0}}$ induced from the identification of ${\Sigma}_{{t}_{0}}$,  ${\Sigma}_{{t}_{1}}$.
\end{note}

The formulae (\ref{spinorHamiltonian})-(\ref{thirdorder}) contain a wealth of information. They are
interpreted in the following remarks.

\begin{remark}
{\bf Commutativity.} The vanishing of the zeroth order (\ref{zerothorder}) just expresses the commutativity of the algebra
of functions. This is a particularity typical of classical geometry and is, of course, not to be expected to hold in
generalizations to noncommutative geometry.
\end{remark}

\begin{remark}
{\bf First order condition}. The vanishing of the first order (\ref{firstorder}) can be traced back to the fact that the spinor field obeys a first
order differential equation. It can be restated in the following formula valid for any Hamiltonian originating from an
arbitrary choice of slicing and time:
\begin{align}
[[ f({t}_{0}), iH], g({t}_{0})] &= 0 \label{orderonecondition}
\end{align}
\end{remark}

\begin{remark}
{\bf Conformal structure}. The second order (\ref{secondorder})
provides the generators $[{\not\! e}_{i},{\not\! e}_{k}]$ of
spacial orthonormal rotations, also called Coriolis fields. Thus,
while it does not provide a length scale, it allows to compare
ratios of lengths in different spacial directions. This means that
the second order fixes the conformal structure. This term is
absent in 1+1 dimensions as there is no conformal information to
talk about.
\end{remark}

\begin{remark}\label{timevectorandscale}
{\bf Time vector and scale}. In 1+1 dimensions, the third order provides the time vector ${\not\! e}_{\bot}$.
This is uniquely extractable by
requiring that it be normalized,
\begin{align}
{{\not\! e}_{\bot}}^{2} &= -1
\end{align}
Moreover, the term sets a scale for the metric, the inverse mass
$\frac {1} {m}$ of the Dirac field. This scale could not be
extracted, e.g., from the spectrum of the Hamiltonian, since an
arbitrary rescaling of time by a constant factor will give a
rescaling of the Hamiltonian. In this sense the Hamiltonian alone
is not sufficient to set a scale.

However, in the general case of $n+1$ dimensions, it is more
useful to abandon the expansion to orders higher than one in time
(For high energy and low energy expansions, see
\cite{Singh-Papini}) and rather {\em postulate} the existence of
a suitable time vector ${\not\! e}_{\bot}$ with the correct
properties. This allows to avoid a complicated extraction
procedure for ${\not\! e}_{\bot}$ and $m$. The ADM-data for
spacetime can then be extracted from
\begin{align}
[[iH,{\not\! e}_{\bot} ],f]&\rightarrow N{g}^{ij}{\not\! e}_{j}\\
tr(iH{\not\! e}_{\bot})&\rightarrow Nm\\
tr([iH,f])&\rightarrow {N}^{i}
\end{align}
This is the approach taken in the next section.
\end{remark}

\begin{remark}
{\bf Gauge and spacial Clifford generators}. The use of the
commutator (\ref{commutatorwithH}) is the following: Given the
time vector, it is easy to recognize the commutator's two terms in
an operationalistic way and to extract thus the shift ${N}^{i}$,
the spacial Clifford generators ${\not\! e}_{\bot}$ together with
their correspondence with spacial codirections. The lapse $N$ is
obtained by checking the prefactor of the first term against
(\ref{thirdorder}). Thus the commutator (\ref{commutatorwithH})
expresses the arbitrary choices (gauge freedom) of the
ADM-formalism, including the choice of a spinor basis and is in
this sense a somewhat supplementary structure.
\end{remark}

The following corollary and remark show that in $1+1$ dimensions the idea of reconstructing spacetime from commutators of
the algebras ${A}_{t}$ meets complete fulfillment (massive case) as well as complete failure (massless case). Both are due
to special properties of the dimension.

\begin{corollary}\label{c1}
In 1+1 dimensions and under the assumption that $m\neq 0$, the geometric structure of spacetime can be recovered from the knowledge of the commutators (\ref{commutatorwithH})and
(\ref{firstorder})-(\ref{thirdorder})
\end{corollary}

\begin{remark}
If $m=0$ in $1+1$ dimensions then the commutator $[f({t}_{1}),g({t}_{0})]$ vanishes at all times. Any (generalized)
eigenspaces of an algebra ${A}_{{t}_{0}}$ can be split into a right-moving eigen-vector and a left-moving eigenvector.
These form a common eigenbasis for all algebras ${A}_{\bullet}$.
\end{remark}


\section{The spectral quadruple}\label{spectralspacetime}

In this section, the structures and properties that were found
useful in Section \ref{globalspacetime} to recover a spacetime
manifold from spectral data are pronounced to be first
principles, whose collection is called a spectral quadruple. At
the same time, they are fused with an imposed symmetry, though it
is hoped that the general suggestions are still visible. The
imposition of symmetry is done for two reasons: First, it
provides a computationally manageable example and in direct
generalization the possibility of a large set of important
analogous examples. Second, the imposed  symmetries provide
automatic smoothness which allows to leave this further
difficulty for this work on the side lines. (For a physically
motivated principle of smoothness see \cite{Kopf1999}.)

Given a category (a collection of objects and morphisms between objects that can be composed, satisfying a certain set of
axioms), a subset of morphisms such that each of its elements has an inverse morphism forms a groupoid.

The following Definition \ref{defspectquad} is to be understood as
a conceptual step and does not attempt to be free of any
redundance. Note that, in the notation of the definition, the
dependence of the operators ${\not\! e}_{\bot}={\not\!
e}_{\bot}(\bullet)$, $\gamma=\gamma (\bullet)$ on the
symbolically written index $\bullet$ is present but mostly
suppressed.

\begin{definition}\label{defspectquad}
The {\bf spectral quadruple} $({A}_{\bullet}, \mathcal{H}, G, C, \gamma (\bullet) ,{\not\! e}_{\bot}(\bullet))$ consists of a collection of
algebras ${A}_{\bullet}$
represented on the Hilbert space $\mathcal{H}$, of a groupoid $G$ and of an antilinear operator $C$.  In addition, for each
of the algebras ${A}_{\bullet}$ two operators ${\not\! e}_{\bot}$, $\gamma$ are given. These structures
satisfy the following conditions:
\begin{enumerate}
\item {\bf Evolution}.

Any two algebras ${A}_{0}$, ${A}_{1}$ of the collection ${A}_{\bullet}$
are required to be mutually unitarily equivalent through a (not
necessarily unique) unitary $U({A}_{0}, {A}_{1})$ and {\em not} mutually commutative,
$[{A}_{{t}_{1}},{A}_{{t}_{2}\neq {t}_{1}}]\neq 0$.
The groupoid $G$ consists of a subset of all
possible unitary equivalences between the algebras in the collection ${A}_{\bullet}$. It is assumed that for
each algebra ${A}_{0}$ in the collection there exists an evolution, a (not necessarily unique) differentiable
path ${\alpha}_{t}:t\in\mathbb{R}\rightarrow {U}_{t}({A}_{0}, {A}_{t})$
with ${\alpha}_{0}= 1$ such that the generator (derivative) at $t=0$, denoted by
$iH$ is compatible with the further requirements.

\item {\bf Charge conjugation.} The antilinear operator $C$ commutes with $G$ and satisfies
\begin{align}
{C}^{2} &= {(-1)}^{s(n)}\\
\intertext{for the spacetime dimension $n$ and with}
s(n)&:=\frac {1} {8} (n-1)(n-2)(n-3)(n-4). 
\end{align}

\item {\bf First order condition (Dynamics)}.
\begin{align}
\left[ \left[ f, iH\right] , {g}^{op}\right] &=0& \text{for any $f,g\in{A}_{0}$ and any generator $iH$,}
\end{align}
with ${g}^{op}=C{g}^{\ast}C$.

\item {\bf The time vector}. For each algebra ${A}_{0}$ in the collection ${A}_{\bullet}$ there exists an operator ${\not\!
e}_{\bot}$ called the time vector satisfying
\begin{align}
{\not\! e}_{\bot}^{2} &= -1\\
{\not\! e}_{\bot}^{\ast}&=-{\not\! e}_{\bot}
\end{align}
and the compatibility conditions in 5. and 6. of this definition.

\item {\bf The volume element}. For any ${A}_{t}$ there exists an operator $\gamma$ such that
\begin{align}
{\gamma}^{2}&=\pm 1&\\
\{ {\not\! e}_{\bot},\gamma \} &= 0 &\text{for even spacetime dimension}\\
[ {\not\! e}_{\bot},\gamma ] &= 0 &\text{for odd spacetime
dimension}
\end{align}
and
\begin{align}
 \gamma &= {\not\! e}_{\bot}
\sum_{{f}_{\bullet}\in {A}_{t}}{{f}_{i_0}[D,{f}_{i_1}]\ldots
[D,{f}_{i_n}]}
&\text{for even spacetime dimension $n+1$}\label{hoch1}\\
\gamma &= \sum_{{f}_{\bullet}\in {A}_{t}}{{f}_{i_0}[D,{f}_{i_1}]\ldots [D,{f}_{i_n}]} \label{hoch2}
&\text{for odd spacetime dimension $n+1$}
\end{align}
\begin{align}
\intertext{and for suitable functions ${f}_{\bullet}$ where $D$ is given by}
D&=
\begin{cases}
\gamma [iH,\gamma ]&\text{for even spacetime dimension}\\
iH&\text{for odd spacetime dimension}\\
\end{cases}&
\end{align}

\item {\bf Geometry of space.} For any algebra ${A}_{t}$ of the collection
$({A}_{t}, \mathcal{H} ,{\not\! e}_{\bot}[H,{\not\! e}_{\bot}], \gamma , C)$ is
\begin{itemize}
\item a spectral triple for odd spacetime dimension.

\item a spectral triple for even spacetime dimension, if restricted to each of the two eigenspaces of ${\not\! e}_{\bot}$.
\end{itemize}
\end{enumerate}
\end{definition}

\begin{remark}
The right hand sides of equations (\ref{hoch1}), (\ref{hoch2})
are to be understood as (images of) Hochschild cycles. The
principal part of the operator $D$ (compare with
(\ref{spinorHamiltonian})) is the same as the principal part of
the spacial Dirac operator but multiplied from the left by
${\not\! e}_{\bot}$. Since ${\not\! e}_{\bot}^{2} = -1$, this has
no effect on the form of (\ref{hoch2}) but requires the
appearance of ${\not\! e}_{\bot}$ in (\ref{hoch2}), to cancel one
superfluous ${\not\! e}_{\bot}$ coming from the commutators. Note
that, together with the requirements in parts 5, 6 of the
definition, it follows that
\begin{align}
D\gamma &= (-1)^{n+1}\gamma D
\end{align}
\end{remark}

\begin{remark}\label{timechoices}
A minimal version of $G$ would be $G=\mathbb{R}$.

An interesting model in this respect would be the classical evolution
induced from the modular automorphism group determined  by a quasi-free state on the Dirac-quantized phase space $H$.
This would make contact with the work of A.Connes and C. Rovelli \cite{Connes-Rovelli}.

However, $G$ may be much larger and it is in this context that the concept of many-fingered time of general relativity
comes to its full expression.

For a large $G$, the definition does not exclude the possibility that, along with generators $iH$ satisfying the
definition, there may be some that may not satisfy the definition. These are to be thought of as in some sense singular and
are to be avoided (see Remark \ref{symtimechoices}).
\end{remark}

\begin{remark}
The last requirement in the definition of a spectral quadruple
puts things on the safe side in order to be sure that reasonable spacetime
geometries deserving to be called manifolds will be obtained .  It is to be
noted though that ${\not\! e}_{\bot}[H,{\not\! e}_{\bot}]$ is in general not
the spacial Dirac operator, though it can be used as such for making sure the
corresponding spacial section is a manifold, since only the principal symbol of
the Dirac operator matters and is in this way obtained correctly.
\end{remark}

\begin{remark}\label{r2}
{\bf Spacetime points}. Compared with the situation in the previous section, the definition of the spectral quadruple provides for more general
situations, in particular for a many-fingered time. This leads to an additional difficulty: Assume, that the algebras
in the collection are commutative. Then it is fine to take the characters of an algebra as being points of spacetimes.
However, some characters of different algebras in the collection describe the same point. In order to compare
characters, one has to extend them as functionals on a larger algebra encompassing all of the collection
${A}_{\bullet}$. But since the characters are distributions, they cannot be extended to all bounded operators on
$\mathcal{H}$.
It is natural to expect that this algebra should be determined by a smoothness principle common to the whole collection
${A}_{\bullet}$.
\end{remark}

Two types of spectral quadruples are of particular interest:

\begin{definition}
A {\bf general spectral quadruple} is a spectral quadruple where $G$ is given by
{\em all} smooth unitary
equivalences  $U({A}_{\bullet}, {A}_{\bullet})$

A {\bf symmetric spectral quadruple} is a spectral quadruple distinguished by
the following
conditions
\begin{enumerate}
\item $G$ a finite dimensional Lie group
(and $iH$ is then in the Lie algebra of $G$).
\item For any algebra ${A}_{{t}_{0}}$, the subgroup $K$ preserving the algebra
coincides with a maximal compact subgroup of $G$.
\item The operators ${\not\! e}_{\bot}$, $\gamma$ commute with the group $K$:
\begin{align}
[k,{\not\! e}_{\bot}]&= 0,\\
[k,\gamma]&= 0,
\end{align}
for any $k\in K$.
\end{enumerate}
\end{definition}

\begin{remark}\label{symtimechoices}
Specializing Remark \ref{timechoices} to a symmetric spectral quadruple, the generator $iH$ for an algebra ${A}_{0}$ is not
to be chosen as a compact generator in the Lie algebra of $G$ preserving ${A}_{0}$ as it may not fulfill the
the geometry-of-space requirement in Definition \ref{defspectquad} of a spectral quadruple.
\end{remark}

Following these general consideration, we will choose a particular class of symmetric spectral quadruples which we will
call {\bf de Sitter spectral quadruples}.
\begin{itemize}
\item The group $G$ is chosen to be ${SL}_{2}(\mathbb{R})$.
\item The set of algebras is generated from one algebra by the action of $G$. This algebra in turn is generated from
one unitary generator $u$:
\begin{align}
{u}^{\ast} = {u}^{-1}
\end{align}
\item The mutual position of $G$ and $u$ on the representation space $\mathcal{H}$ is partially determined by the
commutation relation of $u$, ${\not\! e}_{\bot}$ and $\gamma$ with a compact generator ${T}_{21}$ of the Lie algebra of $G$:
\begin{align}
[{T}_{21},u]&=iu,\label{t21ucommutator}\\
[{T}_{21},{\not\! e}_{\bot}]&= 0,\\
[{T}_{21},\gamma]&= 0.
\end{align}
\item The above set of structures is irreducibly represented. This requirement is to avoid dealing with multiple copies.
\end{itemize}

\begin{note}
The compact generators of a Lie group generate the maximal
compact subgroup of a Lie group \cite{Barut-Raczka}. In the case
of ${SL}_{2}(\mathbb{R})$ a compact subalgebra of the Lie algebra
is just 1-dimensional and thus spanned by one element ${T}_{21}$.
This indicates how to proceed in higher dimensional cases.
\end{note}

As is clear from the definitions, the Hilbert space $\mathcal{H}$ of a de Sitter spectral quadruple is in
particular the representation space of a unitary representation of $G={SL}_{2}(\mathbb{R})$. Therefore, in the next
section, the representation theory of $G={SL}_{2}(\mathbb{R})$ will be reviewed.


\section{${SL}_{2}(\mathbb{R})$ and its representation theory}\label{SL2Rreptheory}

The group ${SL}_{2}(\mathbb{R})$ is treated in a number of sources. The present review is based on
\cite{Barut-Raczka,Lang,Limic-Niederle-Raczka,Ruhl,Robert}.

\begin{definition}
The group
${SL}_{2}(\mathbb{R})$ is the Lie group of $2\times 2$-matrices of real numbers with unit determinant.\\
${sl}_{2}(\mathbb{R})$ is the Lie algebra of traceless $2\times 2$-matrices of real numbers.
\end{definition}

\begin{remark}
(see \cite{Barut-Raczka}). The Lie group ${SL}_{2}(\mathbb{R})$ and the Lie algebra ${sl}_{2}(\mathbb{R})$ have the
following decomposition (Iwasawa decomposition).
\begin{align*}
{SL}_{2}(\mathbb{R}) &=K\times A \times N =
\underbrace{\begin{pmatrix}  \cos{\phi}&\sin{\phi} \\ -\sin{\phi} & \cos{\phi} \end{pmatrix}}_{\text{maximal
compact}}
\underbrace{\begin{pmatrix}  a & 0 \\ 0 & {a}^{-1} \end{pmatrix}
\begin{pmatrix}  1&b \\ 0 & 1 \end{pmatrix}}_{\text{solvable}}\\
{sl}_{2}(\mathbb{R}) &=k \times a \times n =
\underbrace{\begin{pmatrix}  0&c \\ -c & 0 \end{pmatrix}}_{\text{antisymmetric}}
\underbrace{\begin{pmatrix}  d & 0 \\ 0 & -d \end{pmatrix}\begin{pmatrix}  0 & n \\ 0 & 0
\end{pmatrix}}_{\text{solvable}}\\
&\text{for $\phi ,a,b,c,d,n\in \mathbb{R}$}
\end{align*}

${SL}_{2}(\mathbb{R})$  is a simple Lie group and has thus a non-degenerate Killing metric on its Lie
algebra with signature $(-,+,+)$.

${SL}_{2}(\mathbb{R})$ is a double cover for $SO(1,2)$.
\end{remark}

The Lie algebra ${sl}_{2}(\mathbb{R})$ can be given by three generators ${T}_{01}$, ${T}_{02}$ and ${T}_{21}$
satisfying the following commutation relations:
\begin{align}
[{T}_{01},{T}_{21}]&=-{T}_{02}&
[{T}_{02},{T}_{21}]&=+{T}_{01}&
[{T}_{01},{T}_{02}]&=-{T}_{21}
\end{align}

In a complex representation, raising and lowering operators may be defined:
\begin{align}
{T}_{\pm}&= {T}_{01}\mp i{T}_{02}
\end{align}
Together with ${T}_{21}$, these operators generate the complexification of ${sl}_{2}(\mathbb{R})$. Their commutation
relations are:
\begin{align}
[{T}_{21},{T}_{\pm}]&=\pm i {T}_{\pm}\\
[{T}_{+},{T}_{-}]&=-2i {T}_{21}\label{normalization}
\end{align}
In addition, the following holds in unitary representations:
\begin{align}
{T}_{21}^{\ast}&=-{T}_{21}\\
{T}_{01}^{\ast}&=-{T}_{01}\\
{T}_{02}^{\ast}&=-{T}_{02}\\
{T}_{\pm}^{\ast}&=-{T}_{\mp}\label{unitarity}
\end{align}

\subsection*{Representations}

The group ${SL}_{2}(\mathbb{R})$ is a noncompact group. Its unitary
representations are thus bound to be infinite dimensional while there are
non-unitary finite dimensional representations, e.g. the defining
representation.

Any representation can be decomposed into irreducible representations of the
maximal compact subgroup $K$ which is just the circle group ${U}_{1}$. The irreducible
representations of ${U}_{1}$ are 1-dimensional and given by ${\pi}_{n}:\phi\rightarrow
{e}^{in\phi}$ with $n\in\mathbb{Z}$. The maximal compact subgroup can be taken to be
generated by ${T}_{21}$ . That the eigenvalues of ${T}_{21}$
are to be integers or half-integers is a global requirement of the group representation and can in this case not be seen
on the level of the generator ${T}_{21}$ alone but comes from the requirement that the representation of $K={U}_{1}$ should
be single-valued.

We will consider only admissible unitary representations \cite{Lang} in which each eigen-spaces of
${T}_{21}$ are only finite dimensional.

The eigen-vectors of ${T}_{21}$ will be labeled by $\mid
n\rangle$:
\begin{align}
{T}_{21}\mid n\rangle&=in\mid n\rangle&\text{for $n$ integer or half-integer}\\
\intertext{and will be assumed to be normalized}
\langle n\mid n\rangle &= 1. &
\end{align}

For any eigen-vector in such an irreducible representation, the vectors
${T}_{-}{T}_{+}\mid n \rangle$, ${T}_{-}{T}_{+}\mid n \rangle$ are
proportional to $\mid n\rangle$.

Define:
\begin{align}
{T}_{+}\mid n \rangle &= {c}_{n}\mid n+1 \rangle
\intertext{It follows from unitarity (\ref{unitarity}):}
{T}_{-}\mid n+1 \rangle &= -{\bar{c}}_{n}\mid n \rangle
\end{align}
From (\ref{unitarity})  and  (\ref{normalization}), it follows that
\begin{align}
\langle n \mid [{T}_{+},{T}_{-}]n\rangle&=-2i \langle n \mid{T}_{21}n\rangle\\
-\langle {T}_{-}n \mid {T}_{-}n\rangle+\langle {T}_{+}n \mid {T}_{+}n\rangle&=2n\\
{\mid {c}_{n}\mid}^{2} -{\mid {c}_{n-1}\mid}^{2} &= 2n \\
\intertext{and solving the recursion relation gives}
{\mid {c}_{n}\mid}^{2}= {\left( n+\frac {1} {2}\right)}^{2}+{R}^{2}{m}^{2} ,
\end{align}
with ${R}^{2}{m}^{2}$ being a real number (possibly negative! Writing for this number the expression
${R}^{2}{m}^{2}$ is convenient for latter interpretation and has no importance here).
Since ${\mid {c}_{n}\mid}^{2}$ is positive, ${R}^{2}{m}^{2}$ has
to satisfy the following condition
\begin{align}
{\left( n+\frac {1} {2}\right)}^{2}+{R}^{2}{m}^{2}\geq 0\label{repcondition}
\end{align}
for all half-integers $n$ appearing in the representation. This condition allows us to get insight into the
classification of the unitary irreducible representations of ${SL}_{2}(\mathbb{R})$:
\begin{itemize}
\item
As long as
${R}^{2}{m}^{2}>0$, the condition (\ref{repcondition}) is always satisfied and ${c}_{n}$ never
vanishes. Given one eigenvector $\mid n\rangle$ of the hermitean operator $-i{T}_{21}$ and for some half-integer $n$, one obtains eigenvectors $\mid
n+k\rangle$, for all $k\in\mathbb{R}$. These then span the representation space of the irreducible
representation. These representations, label by the number ${R}^{2}{m}^{2}$ and by wheter the eigenvalues
are integers or true halfintegers (i.e., integers + $\frac 1 2$), are referred to as the {\em principal
series}.
\item
If $0\geq {R}^{2}{m}^{2}>\frac {1} {4}$ then the condition (\ref{repcondition}) is always satisfied and
${c}_{n}$ never vanishes under the assumption that the eigenvalues are integers. These representations are
called the {\em complementary series}.
\item
If the condition (\ref{repcondition}) cannot be satisfied for eigenvalues shifted by an arbitrary amount,
then some ${c}_{n}$ have to vanish in order not to run into contradictions with (\ref{repcondition}) while
applying raising and lowering operators. This bounds the eigenvalues of $-i{T}_{21}$ in the unitary irreducible
representation from above or from below. The number ${R}^{2}{m}^{2}$ has then to satisfy
\begin{align}
{R}^{2}{m}^{2}=-{\left( {n}_{0}+\frac {1} {2}\right)}^{2}
\end{align}
for some halfinteger ${n}_{0}$. These unitary irreducible representations are thus discretely indexed by
${n}_{0}$ and by whether the eigenvalues of $-i{T}_{21}$ are bound from above or from below. This set of unitary irreducible
representations is the {\em discrete series}.
\end{itemize}

\section{The symmetric spectral quadruples of de Sitter space}\label{deSitterrecover}

The Hilbert space of the spectral quadruple is required to carry
a representation of the group ${SL}_{2}(\mathbb{R})$ and has in
addition to accommodate the operators $u$, ${\not\! e}_{\bot}$ and
$\gamma$. Since  ${\not\! e}_{\bot}$ and $\gamma$ commute with
${T}_{21}$, they can be considered separately on each eigenspace
of ${T}_{21}$. Mutually, ${\not\! e}_{\bot}$ and $\gamma$
anticommute and do not vanish, as follows from the definition of
the spectral quadruple. Thus they force the eigenspaces
${\mathcal{H}}_{n}$ of ${T}_{21}$ to be at least 2-dimensional.
In addition, $u$ is an invertible raising operator and its
inverse ${u}^{\ast}$ an invertible lowering operator, as follows
from Equation (\ref{t21ucommutator}). Thus all the eigenspaces
${\mathcal{H}}_{n}$ of ${T}_{21}$ are bound to have the same
dimension. From the requirement of irreducibility and from the
invertibility of the operator $u$ we obtain that this dimension
has to be 2 and that the discrete series can therefore not be a
candidate.

We will further assume that the eigenvalues are true half-integers, as we are interested in
interpreting the representation space as the phase space of a spinor field. This excludes the
complementary series of representations as well as the part of the principal series with
integer eigenvalues of $-i{T}_{21}$.

A common orthonormal eigenbasis $\mid n,\pm\rangle$ for the mutually commuting antihermitean operators
${T}_{21}$,  ${\not\!e}_{0}$ can be given:
\begin{align}
{T}_{21}\mid n,\pm\rangle &= in\mid n,\pm\rangle\\
{\not\!e}_{0}\mid n,\pm\rangle &= \pm i\mid n,\pm\rangle\\
\langle {n}^{'}, {sign}^{'}\mid n,sign\rangle &= {\delta}_{{n}^{'}n}{\delta}_{{sign}^{'}sign}
\end{align}
The eigenvectors are then unique, up to phases. The relative phases are fixed by requiring
further
\begin{align}
u\mid n,\pm\rangle &= \mid n+1,\pm\rangle \label{u}
\end{align}
This leaves only two phases free, one of them an overall phase.


\subsection*{The order-one condition}

The only unspecified operators of the spectral quadruple are now
${T}_{+}$ and ${T}_{-}$, the raising and lowering operators.

They are given by their restrictions
\begin{align}
{T}_{\pm}(n):{\mathcal{H}}_{n}\rightarrow {\mathcal{H}}_{n\pm 1}
\end{align}

The eigenspaces ${\mathcal{H}}_{n}$ can be identified using the
unitary bijections given by restrictions of $u$. Then the
operators ${T}_{\pm}(n)$ can all be understood to operate on a
common, 2-dimensional Hilbert space.

The order-one condition for the raising and lowering operators
can then be written as
\begin{align}\label{recursion}
{T}_{\pm}(n+1)-2{T}_{\pm}(n)+{T}_{\pm}(n-1)= 0
\end{align}
Note that
\begin{align}
{{T}_{+}(n)}^{\ast}&=-{T}_{-}(n+1)\\
{{T}_{-}(n)}^{\ast}&=-{T}_{+}(n-1)
\end{align}
so that both equations (\ref{recursion}) can be solved recursively, if two restrictions
${T}_{\pm}(\bullet)$ are known.

 In order to keep the symmetric
way in which the raising and lowering operators appear in the
calculations, we choose as the two required restrictions to be
${T}_{+}(+\frac {1} {2})$, ${T}_{-}(-\frac {1} {2})$.

From the previous section, it is clear that the ${T}_{\pm}(n)$
have to be unitaries times the factor $\sqrt{{\left( n+\frac {1}
{2}\right)}^{2}+{R}^{2}{m}^{2}}$.

We set
\begin{align}
{T}_{+}(+\frac {1} {2})&=\sqrt{1+{R}^{2}{m}^{2}}\,\, {u}_{+}&{u}_{+}^{\ast}{u}_{+}&={u}_{+}{u}_{+}^{\ast}=\mathbf{1}\\
{T}_{-}(-\frac {1} {2})&=\sqrt{1+{R}^{2}{m}^{2}}\,\, {u}_{-}&{u}_{-}^{\ast}{u}_{-}&={u}_{-}{u}_{-}^{\ast}=\mathbf{1}
\end{align}

Solving the
order-one recursion condition, one obtains
\begin{align}
{T}_{+}(n)&= \sqrt{1+{R}^{2}{m}^{2}}
\left( \left(+\frac {n} {2}- \frac {1} {4}\right)({u}_{+}+{u}_{-}^{\ast}) + {u}_{+}\right)\\
{T}_{-}(n)&= \sqrt{1+{R}^{2}{m}^{2}}
\left( \left(-\frac {n} {2}- \frac {1} {4}\right)({u}_{-}+{u}_{+}^{\ast}) + {u}_{-}\right)
\end{align}

\subsection*{Charge conjugation.}

It remains to determine the $2\times 2$-matrices ${u}_{+}$, ${u}_{-}$.

For that purpose we use a general parametrization of unitary $2\times 2$-matrices,
\begin{align}
{u}_{+}&= {e}^{i{\rho}_{+}}\frac {1} {\sqrt{1+{x}_{+}^{2}+{y}_{+}^{2}}}
\begin{pmatrix}
-i{x}_{+} +\tanh{{\theta}_{+}}            & \frac {1} {\cosh{{\theta}_{+}}}+i{y}_{+}\\
-\frac {1} {\cosh{{\theta}_{+}}}+i{y}_{+}   & i{x}_{+} +\tanh{{\theta}_{+}}
\end{pmatrix}\\
{u}_{-}&={e}^{i{\rho}_{-}}\frac {1} {\sqrt{1+{x}_{-}^{2}+{y}_{-}^{2}}}
\begin{pmatrix}
-i{x}_{-} +\tanh{{\theta}_{-}}            & \frac {1} {\cosh{{\theta}_{-}}}+i{y}_{-}\\
-\frac {1} {\cosh{{\theta}_{-}}}+i{y}_{-}   & i{x}_{-} +\tanh{{\theta}_{-}}
\end{pmatrix}
\end{align}

As a consequence of the commutation of the charge conjugation $C$ with $G$ in the definition of a spectral quadruple, one
obtains
\begin{align}
{e}^{i{\rho}_{}}:={e}^{i{\rho}_{+}}&=-{e}^{i{\rho}_{-}}\\
{x}_{}:={x}_{+}&=-{x}_{-}\\
{\theta}_{}:={\theta}_{+}&={\theta}_{-}\\
{y}_{}:={y}_{+}&={y}_{-}
\end{align}

Out of the remaining four parameters, two of them, $\rho$ and $y$
can be set to zero, since they exactly correspond to the two
remaining phases in the above basis choice. $\rho$ is the overall
phase and $y$ is connected with a phase freedom in the spinor
basis as will become apparent once the geometric interpretation
of the de Sitter quadruple is reached.

Thus the de Sitter spectral quadruples form a two-parametric family. Comparing with the representation theory
of ${SL}_{2}(\mathbb{R})$, one finds that $x$ can be identified with the parameter $Rm$ while there are no further
restrictions on $\theta$. Thus the two matrices ${u}_{+}$ and ${u}_{-}$ can be written as

\begin{align}
{u}_{+}&=
  \begin{pmatrix}
  -iRm + \tanh{\theta}       & \frac {1} {\cosh{\theta}}\\
  -\frac {1} {\cosh{\theta}} & iRm + \tanh{\theta}
  \end{pmatrix},\\
  \intertext{ }
{u}_{-}&=
  \begin{pmatrix}
  -iRm + \tanh{\theta}       & -\frac {1} {\cosh{\theta}}\\
   \frac {1} {\cosh{\theta}} & iRm + \tanh{\theta}
  \end{pmatrix}.
\end{align}

Following Remark \ref{timevectorandscale} and Remark \ref{r2} (or Corollary \ref{c1} and Remark \ref{r2}, since
this is the special, 1+1 dimensional case), the corresponding spacetime geometry can be
directly calculated.

The result is 1+1 dimensional de Sitter space of radius $R$ containing a Dirac spinor field of mass
$m$. The mass $m$ and  the radius $R$ may be reinterpreted at will, as long as $Rm$ is kept constant. In that sense, the
mass of the Dirac field becomes the meter stick. The parameter does not change the geometry: The
algebra generated by $u$, from which all other algebras are obtained by symmetry group actions describes a spacelike circle
invariant under the action of ${T}_{21}$. There is a one parameter set of such circles. $\theta$ can be chosen as that
parameter.

Instead of carrying out the detailed calculations, one can get these results by checking that the de Sitter spectral
quadruples obtained here are identical with the spectral data of 1+1 de Sitter space calculated in the Appendix, see
(\ref{sq1})-(\ref{sq6}).

\begin{remark}
The fact that different symmetric spectral quadruples describe the same geometry is not surprizing. It is a consequence of
$G$ having been chosen too small to provide for all unitary equivalences that take any smooth spacelike circle into any other
smooth spacelike circle on de Sitter. A one-to-one correspondence between
spacetimes and spectral quadruples should be searched for with a sufficiently large choice of $G$ only.
\end{remark}

\begin{remark}
Going through a direct calculation, one cannot avoid the problem
of extending characters of one algebra in the collection to
generalized vectors, on which other algebras may act. This
requires to see the characters as distributions (functionals) on
a dense subspace of the Hilbert space \cite{Gelfand-Shilov3}. But
suitable subspaces of test vectors are in this case obtained for
free from the action of the group $G$: They can be chosen to
consist of  vectors with rapidly decaying coordinates in the
eigen-basis of  ${T}_{21}$. In this point, the employed Lie
symmetry provides substantial help with matters of smoothness.
\end{remark}
\section{Finite spectral quadruples}\label{discretequadruples}

A finite spectral quadruple is based on a collection of finite
dimensional algebras with the necessary additional structures.

Here, we will build a spectral
quadruple out of a spectral triple $({A}_{0},\mathcal{H} ,D,\gamma ,J)$. This
allows us to take advantage of the
fact that the relevant spectral triples have already been fully characterized
\cite{Paschke-Sitarz1998,Krajewski1998}: Finite spectral triples are classified
by their algebra $A$ which is a finite direct
sum of matrix algebras, $A=\oplus_{i=1}^{k} {M}_{{n}_{i}}(\mathbb{C})$ and a
symmetric and invertible $k\times k$-matrix with integer entries, the intersection form
${q}_{ij}:{K}_{\bullet}(A)\times {K}_{\bullet}(A)\rightarrow \mathbb{Z}$, from
which the remaining data of the spectral
quadruple can be read off. In particular, the Hilbert space $\mathcal{H}$ is
given by
\begin{align}
\mathcal{H}=\bigoplus_{i,j}^{k}\mathcal{H}_{ij}:=
\bigoplus_{i,j}^{k}({\mathbb{C}}^{{n}_{i}}\otimes
{\mathbb{C}}^{\mid {q}_{ij}\mid } \otimes {\mathbb{C}}^{{n}_{j}}),
\end{align}
on which the algebra $A$ acts by left multiplication as
$a\otimes\mathbf{1}\otimes\mathbf{1}$ for $a \in A$ while the opposite algebra
$J{A}^{\ast}J$, with $J:{H}_{ij}\rightarrow{H}_{ji}$,
$J({v}_{i}\otimes{v}_{ij}\otimes{v}_{j})={\bar{v}}_{j}\otimes{\bar{v}}_{ij}\otimes{\bar{v}}_{i}$
antilinear, acts by right matrix multiplication as
$\mathbf{1}\otimes\mathbf{1}\otimes {a}^{T}$ for $a \in A$.
Furthermore,
\begin{align}
\gamma \mathcal{H}_{ij} &=\mathcal{H}_{ij} &
\gamma{\mid}_{\mathcal{H}_{ij}}&=sign({q}_{ij})\mathbf{1}_{\mathcal{H}_{ij}}
\intertext{and the Dirac operator
$D:\mathcal{H}_{kl}\rightarrow\mathcal{H}_{ij}$ satisfies the equations}
D_{ij\, kl}&=D_{ij\, kl}^{+} &D_{ij\, kl}&=\overline{D_{ji\, lk}}\\
\intertext{following from the requirements}
D&={D}^{\ast} & JD &= DJ\\
\end{align}
The Dirac operator is further restricted by the order-one condition of a
spectral triple.

\begin{example}\label{twopointexample}
Let a finite spectral triple be given by the algebra
\begin{align}
A&=\mathbb{C}\oplus\mathbb{C}\\
\intertext{and by the intersection form}
q&=
\begin{pmatrix}
1&-1\\
-1&0\\
\end{pmatrix}.\\
\intertext{Then}
\mathcal{H} &= {\mathbb{C}}^{3}\\
\gamma&=
\begin{pmatrix}
1&0&0\\
0&-1&0\\
0&0&-1
\end{pmatrix},\\
J&= \text{complex conjugation} \times
\begin{pmatrix}
1&0&0\\
0&0&1\\
0&1&0
\end{pmatrix}.
\end{align}

The representations $\pi$, ${\pi}^{op}$ of the algebra $A$ and of the opposite
algebra ${A}^{op}$ are given by
\begin{align}
\pi ({z}_{1}\oplus {z}_{2})=
\begin{pmatrix}
{z}_{1}&0&0\\
0&{z}_{1}&0\\
0&0&{z}_{2}
\end{pmatrix},\\
{\pi}^{op} ({z}_{1}\oplus {z}_{2})=
\begin{pmatrix}
{z}_{1}&0&0\\
0&{z}_{2}&0\\
0&0&{z}_{2}
\end{pmatrix},\\
\intertext{and}
D&=
\begin{pmatrix}
0&m&\bar{m}\\
\bar{m}&0&0\\
m&0&0
\end{pmatrix},\qquad\text{for $m\in\mathbb{C}$}\label{finitediracform}\\
\end{align}
\end{example}

In order to construct a spectral quadruple,
it is now sufficient to find in addition to the spectral triple data an operator
${\not\! e}_{\bot}$:
\begin{align}
{{\not\! e}_{\bot}}^{2}&=-\mathbf{1},\\
{{\not\! e}_{\bot}}^{\ast}&=-{\not\! e}_{\bot},\\
[{\not\! e}_{\bot}, \gamma]&=0,
\end{align}

as required by the definition of a 1+0-dimensional spectral quadruple,
and to supply a Hamiltonian $H$ as a function of time.

The operator ${\not\! e}_{\bot}$ and the Hamiltonian $H$ can be chosen as
\begin{align}
{\not\! e}_{\bot} := i\gamma (t),\label{discretee}\\
H(t):= {\not\! e}_{\bot} (t) D(t).
\end{align}
Denoting the evolution operator by $U(t)$, one can write the time dependent operator $D(t)$ as
${U}^{-1}(t) {D}_{0}(t) U(t)$ with
respect to a comoving basis in which the representation of
the algebra remains diagonal. Here ${D}_{0}(t)$ has the same form for all times $t$ as the Dirac operator  $D$ just with
its entries replaced by smooth
functions .

In particular, ${D}_{0}(t)$
has the form (\ref{finitediracform}), if the spectral triple of Example \ref{twopointexample} is taken.
The equal time distance of the two points ${x}_{t}$, ${y}_{t}$ of this discrete spectral geometry can be calculated via Connes' distance
formula \cite{Connes94}:
\begin{align}
d({x}_{t},{y}_{t})&=\sup_{{a}_{t}\in {A}_{t},\parallel [D(t),{a}_{t}]\parallel \leq 1}
                    {\left\{ \mid {x}_{t}({a}_{t})-{y}_{t}({a}_{t})\mid \right\}}\\
                  &=\sup_{{a}_{0}\in {A}_{0},\parallel [{D}_{0}(t),{a}_{0}]\parallel \leq 1}
                    {\left\{ \mid {x}_{0}({a}_{0})-{y}_{0}({a}_{0})\mid \right\}}=\\
                  &=\frac {1} {\mid m(t) \mid}
\end{align}
Thus this spectral quadruple describes two points with varying spacial distance  $\frac {1} {\mid m(t) \mid}$.

To determine the distance between the points, rather than to
prescribe it, one would have to provide the corresponding theory
of gravity given by an action functional. E.g., one could use a
spectral action for the Wick rotated system, thus avoiding
problems of non-definiteness of a Lorentzian action.
Alternatively, one could try to write down an action directly, in
terms of $H$ and its time derivative $\dot{H}$. However, $H$ is
noninvertible and it becomes difficult to write a term invariant
under time parametrization. For instance, typical terms like
$\frac {{\dot{H}}^{2}} {{H}^{3}}$, $\sqrt[4]{\dot{H}}$ are pested
with singularities while $\frac {\dot{H}} {H}$ leads to the
solution $H=0$. But this is beyond the scope of the present work.

\begin{remark}
The here presented type of a discrete spectral quadruple is built in a rather naive way on the basis of a discrete
spectral triple, just by adding time. Its value is in showing how the Lorentzian signature is taken up in the discrete
case (see especially (\ref{discretee})) rather than in examining all possibilities of the discrete case.
\end{remark}

\section{Conclusion}

In this work, we have arrived at a spectral
characterization of  spin manifolds with Lorentzian metric through a spectral
quadruple.  While the spectral geometry of Riemannian spin manifolds as given
by the spectral triple has motivated and was incorporated into the spectral
quadruple,  the spectral quadruple is not just a stack of spectral triples
with  time vectors  ${\not\! e}_{\bot}$ and some sign changes added on. The
results are derived from the Hamiltonian rather than from the Dirac operator,
allowing thus for the many-fingered time typical of general relativity. The
requirements of the order-one condition and of charge conjugation are
significantly generalized with the first of the two containing the main
information on dynamics: These requirements are not to hold for one algebra but
instantly for all algebras in the collection ${A}_{\bullet}$ at once.

Note also that the axioms do not presuppose global hyperbolicity.
The lack of global hyperbolicity and the eventual occurrence of
closed timelike curves leaves one still with a meaningful
structure. E.g., closed timelike curves will not allow an
arbitrary choice of initial data on spacial sections and will thus
severely restrict the size of the phase space but still a fuzzy
description of spacetime is eventually left. This should not be
seen as a failure. It simply means that certain details of
geometry, though possibly existent in mathematical imagination,
are automatically dropped, if there is no field to resolve them.
This is not unrelated to the features of the euclidean spectral
quadruple presented in Joke \ref{jk} at the end of this section.

On the practical side,  the notion of a spectral quadruple  is
backed up by the  nontrivial example of de Sitter spectral
quadruples which successfully describe  $1+1$-dimensional de
Sitter space. This is an achievement in itself as it is  not so
easy, despite the generality of spectral geometry, to produce
examples\footnote{This does not mean that there are no examples
available in the Riemannian case, as the collection has been
built up for some time. But still this is an area in need of
substantial further work.}. While the requirements of the
spectral quadruple are in a preliminary form and confrontation
with further examples is definitely needed, a number of important
examples is under firm control: It is clear, how to deal with
generalizations to higher dimensional de Sitter spaces and other
cosmologies of high symmetries. Situations of such high
symmetries are not just toy examples but have  claim to be
directly interpreted: the Friedmann-Robertson-Walker class of
models is the one used as the background of realistic cosmology.

With this in hand it is now only natural to have a second look at
the standard model of elementary particle physics along the lines
of \cite{Connes-Lott}. Judging from the results of  Section
\ref{discretequadruples}, in particular Equation
(\ref{discretee}) , no fermion doubling is to be expected.

\begin{joke}\label{jk}
It is amusing to note that the machinery of the spectral
quadruple which was designed to deal with Lorentzian manifolds
can be turned onto Euclidean situations. To give an example, we
define a class of symmetric spectral quadruples which we will
call {\bf spherical spectral almost quadruples}. The word {\em
almost} refers to one necessary adaptation of the definition of a
spectral quadruple to the euclidean signature: instead of
${\not\! e}_{\bot}^{2}=-\mathbf{1}$ we require ${\not\!
e}_{\bot}^{2}=+\mathbf{1}$.
\begin{itemize}
\item The group $G$ is chosen to be ${SU}_{2}$.
\item The set of algebras is generated from one algebra by the action of $G$. This algebra in turn is generated from
one unitary generator $u$:
\begin{align}
{u}^{\ast} = {u}^{-1}
\end{align}
\item The mutual position of $G$ and $u$ on the representation space $\mathcal{H}$ is partially determined by the
commutation relation of $u$, ${\not\! e}_{\bot}$ and $\gamma$ with a compact generator ${T}_{21}$ of the Lie algebra of
$G$:
\begin{align}
[{T}_{21},u]&=iu,\label{spheret21ucommutator}\\
[{T}_{21},{\not\! e}_{\bot}]&= 0,\\
[{T}_{21},\gamma]&= 0.
\end{align}
\item The above set of structures is irreducibly represented. This requirement is to avoid dealing with multiple copies.
\end{itemize}
The resulting space should be thought of as being a sphere ${S}^{2}$ with an axis of rotational symmetry determined by
the chosen generator ${T}_{21}$. The algebra generated by $u$ describes a circle preserved under the rotational symmetry
generated by ${T}_{21}$. This circle may or may not be chosen at the equator of the sphere and is moved over the sphere by
the action of the group ${SU}_{2}$. In this way any point of the sphere is reached by some circles. However, points of the
sphere can be resolved only in a limit for the following reason:
In analogy with the de Sitter example, the algebras are represented on the space of solutions of the counterpart of the
Dirac equation
\begin{align}
\dirac \psi &= \left( l+\frac {1} {2}\right) \psi
\end{align}
for a fixed angular momentum $l$ of the spherical harmonics. Since ${SU}_{2}$ is a compact group, the Hilbert space of any
unitary irreducible representation is finite dimensional and so is the Hilbert space of the spectral
quadruple.  As seen from the point of view of a true sphere only maximally
localized vectors in the Hilbert space rather than points of the sphere may be found and the representation of the
algebras generates the matrix algebra on the Hilbert space. Thus the spherical spectral almost quadruples describe
the fuzzy spheres.
\end{joke}

\section{Acknowledgements}
The authors would like to thank Jos\'e M. Gracia-Bond\'{\i}a and
Florian Scheck for their encouragement and for a number of useful
discussions and remarks and Andrzej Sitarz and Joseph C
V\'{a}rilly for a number of comments on earlier drafts. T.K. most
thankfully acknowledges support from the Alexander von Humboldt
Foundation.

\section{Appendix}

In this appendix, a selfcontained exposition of de Sitter
space and of spinor fields on de Sitter space is given. The
main purpose is to collect formulae and facts on this model
for use as a testing case for Lorentzian spectral geometry.
We will concentrate on 2-dimensional de Sitter space.
However, at instances where this restriction does not bring
any simplification, the general $n$-dimensional case is dealt with directly.

In the course of this, an important particularity is
obtained: There is a set of spectral data that can be
compared with the spectral quadruples of Section \ref{deSitterrecover}
to provide for one of the main results of this paper: The de
Sitter spectral quadruples describe de Sitter space.

The first three subsections deal with the differential
geometry, with the symmetries and with  the spin geometry of de Sitter space
from
both the point of view of an imbedding into Minkowski space
as well as from the intrinsic point of view. In the
following two subsections,  the Dirac operator of de Sitter space is discussed
and employed to
construct the action of the symmetries on the space of
solutions of the Dirac equation. There, the spectral data in a form comparable
to the de
Sitter spectral quadruples are obtained.

\subsection{The geometry of de Sitter space}

\begin{definition}\label{deSitterdefinition}
The n-dimensional de Sitter geometry is a submanifold of $n+1$ dimensional
Minkowski space given in a fixed Lorentz frame by the equation
\begin{align}
-{({x}^{0})}^{2}+{({x}^{1})}^{2}+\ldots +{({x}^{n})}^{2}={R}^{2}
\end{align}
with the metric induced from the metric of the Minkowski imbedding space,
\begin{align}
 {ds}^{2}=-{(d{x}^{0})}^{2}+{(d{x}^{1})}^{2}+\ldots +{(d{x}^{n})}^{2}.
\end{align}
The  induced metric has signature $(-,\underbrace{+,+,\ldots ,+,+}_{n-1})$.
\end{definition}

\begin{remark}
The intrinsic de Sitter geometry may be defined without any reference to an
imbedding space. This is not done here as
the imbedding space proves to be a very useful technical tool for doing
calculations. But for the sake of clarity, some
calculations are done or redone in an entirely intrinsic way.
\end{remark}

Following Definition \ref{deSitterdefinition}, the two-dimensional de Sitter
geometry can be embedded into the
three-dimensional Minkowski space as the hyperboloid
\begin{align}
-{({x}^{0})}^{2}+{({x}^{1})}^{2}+{({x}^{2})}^{2}={R}^{2},
\end{align}
and  can be parameterized by generalized spherical coordinates
\begin{align}
{x}^{0}&=R\sinh{\theta}& \theta &\in (-\infty,\infty)\\
{x}^{1}&=R\cosh{\theta}\cos{\phi}& \phi &\in [0,2\pi )\\
{x}^{2}&=R\cosh{\theta}\sin{\phi}& &
\end{align}

The coordinate vectors of the spherical coordinates
$\frac {\partial} {\partial\theta}$ $\frac {\partial} {\partial\phi}$ are
expressed in terms of the cartesian basis
$\frac {\partial} {\partial {x}^{0}}$, $\frac {\partial} {\partial {x}^{1}}$,
$\frac {\partial} {\partial {x}^{2}}$
of Minkowski space by:
\begin{align}
\frac {\partial} {\partial\theta}&= \frac {\partial{x}^{i}} {\partial\theta}
\frac {\partial} {\partial{x}^{1}}=
 R\cosh{\theta} \frac {\partial} {\partial {x}^{0}} +
R \sinh{\theta}\cos{\phi}\frac {\partial} {\partial {x}^{1}} +
R \sinh{\theta}\sin{\phi}\frac {\partial} {\partial {x}^{2}}   \\
\frac {\partial} {\partial\phi}&=\frac {\partial{x}^{i}} {\partial\phi} \frac
{\partial} {\partial{x}^{1}}=
R\cosh{\theta}(-\sin{\phi}\frac {\partial} {\partial {x}^{1}}+
\cos{\phi}\frac {\partial} {\partial {x}^{2}})
\end{align}
The induced metric is then
\begin{align}
{g}_{\theta\theta}&= -{R}^{2} & {g}_{\theta\phi}&={g}_{\phi\theta}=0 &
{g}_{\phi\phi} = {R}^{2}\cosh^{2}{\theta}
\end{align}
\begin{align}
{g}_{\bullet\bullet}&=
\begin{matrix}
&\begin{matrix} \theta & \phi\end{matrix}\\
\begin{matrix}  \theta \\ \phi\end{matrix}&
\begin{pmatrix}  -{R}^{2}&0 \\ 0 & {R}^{2}\cosh^{2}{\theta} \end{pmatrix}
\end{matrix}
&
{g}^{\bullet\bullet}&=
\begin{matrix}
&\begin{matrix} \theta & \phi\end{matrix}\\
\begin{matrix}  \theta \\ \phi\end{matrix}&
\begin{pmatrix}  -\frac {1} {{R}^{2}}&0 \\ 0 & \frac {1}
{{R}^{2}\cosh^{2}{\theta}} \end{pmatrix}
\end{matrix}
\end{align}
The symbol $\bullet$ is used to indicate various types of indices without giving
them names.

The Christoffel symbols of the unique torsion-free metric connection
${\nabla}_{\bullet}$ can be calculated as
\begin{align}
{{\Gamma}^{A}}_{BC} &= \frac {1} {2}
{g}^{AD}({g}_{BD,C}+{g}_{DC,B}-{g}_{BC,D})\\
\intertext{with the non-zero components}
{{\Gamma}^{\theta}}_{\phi\phi} &= \cosh{\theta} \sinh{\theta}\\
{{\Gamma}^{\phi}}_{\theta\phi} &= {{\Gamma}^{\phi}}_{\phi\theta}=
\frac {\sinh{\theta}} {\cosh{\theta}}
\end{align}

The wave operator (Laplace-Beltrami) is then
\begin{align}
{\nabla}^{A}{\nabla}_{A} f &= {g}^{AB}{\nabla}_{A}{\nabla}_{B} f=
{g}^{AB} \left( \frac {\partial} {\partial {\xi}^{A}} \frac {\partial } {\partial {\xi}^{B}} -
{{\Gamma}^{C}}_{AB} \frac {\partial} {\partial {\xi}^{C}}      \right)  f=\\
&=-\frac {1} {{R}^{2}} \left( \frac {1} {\cosh{\theta}} \frac {\partial}
{\partial \theta} \theta \frac {\partial} {\partial \theta}- \frac {1} {\cosh^{2}{\theta}}\frac {{\partial}^{2}} {{\partial \phi}^{2}}\right) f ,
\end{align}

Apart from the above intrinsic structures, the extrinsic curvature
${{K}_{A}}^{B}$ of the imbedding of de Sitter space into Minkowski
space may be calculated, see Lemma \ref{extrinsiccurvaturecalculation}.

\subsection{Symmetries}

The de Sitter spacetime is a homogeneous space with respect to the Lorentz group
of the embedding space. The Lorentz
group is generated by boosts and rotations. The generators can be given
explicitly, decorated for convenience by a
double-index:

\begin{align}
\intertext{rotation with axis ${x}^{0}$}
{L}_{21}&={x}^{1}\frac {\partial} {\partial {x}^{2}} -{x}^{2}\frac {\partial}
{\partial {x}^{1}} =
\frac {\partial} {\partial \phi}\\
\intertext{boost with axis ${x}^{1}$}
{L}_{02}&={x}^{0}\frac {\partial} {\partial {x}^{2}} +{x}^{2}\frac {\partial}
{\partial {x}^{0}} =
\sin{\phi}\frac {\partial} {\partial \theta} + \cos{\phi}\tanh{\theta} \frac
{\partial} {\partial \phi}\\
\intertext{boost with axis ${x}^{2}$}
{L}_{01}&={x}^{0}\frac {\partial} {\partial {x}^{1}} +{x}^{1}\frac {\partial}
{\partial {x}^{0}} =
\cos{\phi}\frac {\partial} {\partial \theta} - \sin{\phi}\tanh{\theta} \frac
{\partial} {\partial \phi}
\end{align}

The generators satisfy the commutation relations
\begin{align}
[{L}_{01},{L}_{21}]&={L}_{02}\label{L1}\\
[{L}_{02},{L}_{21}]&=-{L}_{01}\\
[{L}_{01},{L}_{02}]&={L}_{21}\label{L3}
\end{align}

The Casimir operator is
\begin{align}
{L}^{\bullet} {L}_{\bullet} = -{R}^{2}{\nabla}^{A}{\nabla}_{A}
\end{align}

The action of (the double cover of) these symmetries on spinors and spinor
fields is given in the next section, after a
description of spinors.

\subsection{Spinors on de Sitter space}

It is possible to construct spinors either intrinsically or as seen from the
Minkowski space into which the de Sitter
geometry is embedded.

In both cases it is useful to fix a frame (triad).

For the extrinsic description we chose
\begin{align}
{b}_{i} &= \frac {\partial} {\partial {x}^{i}}
\end{align}

For the intrinsic description, only two vector fields ${e}_{0}$, ${e}_{2}$ are needed but for further calculations it is useful to add a third
vector field ${e}_{1}$, normal to the manifold in embedding space. The resulting triad is then:
\begin{align}
{e}_{0} &= \frac {1} {R} \frac {\partial}{\partial\theta}=
\frac {1} {R} \frac {\partial {x}^{i}}{\partial\theta}\frac {\partial}{\partial
{x}^{i}}\\
{e}_{1} &= \frac {\partial}{\partial R}=\frac {1} {R} {x}^{i}\frac {\partial}
{\partial {x}^{i}}\\
{e}_{2} &= \frac {1} {R\cosh{\theta}} \frac {\partial}{\partial\phi}=
\frac {1} {R\cosh{\theta}} \frac {\partial {x}^{i}}{\partial\phi}\frac
{\partial} {\partial {x}^{i}}
\end{align}

\begin{align}
{e}_{0} &=
\cosh{\theta} \frac {\partial}{\partial {x}^{0}}+
\sinh{\theta}\cos{\phi}\frac {\partial}{\partial {x}^{1}}+
\sinh{\theta}\sin{\phi}\frac {\partial}{\partial {x}^{2}}\\
{e}_{1} &=
\sinh{\theta} \frac {\partial}{\partial {x}^{0}}+
\cosh{\theta}\cos{\phi}\frac {\partial}{\partial {x}^{1}}+
\cosh{\theta}\sin{\phi}\frac {\partial}{\partial {x}^{2}}\\
{e}_{2} &=
-\sin{\phi}\frac {\partial}{\partial {x}^{1}}+
\cos{\phi}\frac {\partial}{\partial {x}^{2}}
\end{align}

The two frames coincide for
\begin{align}
\theta &=0 & \phi = 0.\label{agreeinpoint}
\end{align}

The triad can then be chosen to generate the corresponding Clifford algebra for
the Clifford generators ${\not
\!e}_{\bullet}$ by the anticommutation relations

\begin{align}
 {\not \!e}_{i}{\not \!e}_{j}+{\not \!e}_{j}{\not \!e}_{i}&={g}_{ij} \mathbf{1}
\end{align}

The same anticommutation relations are satisfied by the Clifford generators
corresponding to ${b}_{\bullet}$ but usually denoted
by ${\gamma}_{\bullet}$ rather than by ${\not\!b}_{\bullet}$
One is free to choose a basis in the spinor space and this freedom will be used
here to get the following
representation of the Clifford generators ${\gamma}_{\bullet}$ (and similarly
for ${\not\!e}_{\bullet}$).

\begin{align}
{\gamma}_{0}&=\begin{pmatrix}  i&0 \\ 0 & -i \end{pmatrix}&
{\gamma}_{1}&=\begin{pmatrix}  0&-i \\ i & 0 \end{pmatrix}&
{\gamma}_{2}&=\begin{pmatrix}  0&1 \\ 1 & 0 \end{pmatrix}
\end{align}

The choice was made in such a way as to make ${\gamma}_{0}$ diagonal which seems
to be useful in later computations.

\subsubsection{Intrinsic description}

The vectors ${e}_{0}$, ${e}_{2}$ are inside the de Sitter space
and ${e}_{1}$ is orthogonal to it. ${e}_{1}$ can of course not be
part of an intrinsic description but is useful for matters of
comparison with extrinsic calculations.

The intrinsic spin connection ${D}_{A}$ can be calculated from the connection
form
\begin{align}
{{{\omega}_{A}}^{b}}_{c}&={e}_{B}^{b}{\nabla}_{A}{e}_{c}^{B}=
{e}_{B}^{b}{\partial}_{A}{e}_{c}^{B}+{e}_{C}^{b}{{\Gamma}^{CB}}_{A}{e}_{c}^{B}
\end{align}

The forms
\begin{align}
{e}^{0}&= R d\theta\\
{e}^{2}&= R\cosh{\theta} d\phi
\end{align}
are a basis dual to ${e}_{0}$, ${e}_{2}$.

\begin{align}
\begin{split}
{{{\omega}_{A}}^{0}}_{2}&={e}_{B}^{0}{\nabla}_{A}{e}_{2}^{B}=
\underbrace{{e}_{B}^{0}{\partial}_{A}{e}_{2}^{B}}_{0}+{e}_{C}^{0}{{\Gamma}^{C}}_{AB}{e}_{2}^{B}=\\
&=\frac {1}{\cosh{\theta}}{{\Gamma}^{\theta}}_{A\phi} = \frac
{1}{\cosh{\theta}}d\phi \,{{\Gamma}^{\theta}}_{\phi\phi}=
 \sinh{\theta} d\phi
\end{split}
\end{align}

\begin{align}
{\nabla}_{0}{e}_{0} &= 0&
{\nabla}_{0}{e}_{2} &= 0\\
{\nabla}_{2}{e}_{0} &= \frac {1} {R} \tanh{\theta} {e}_{2}&
{\nabla}_{2}{e}_{2} &= \frac {1} {R} \tanh{\theta} {e}_{0}
\end{align}

From the (tangent space) connection form ${\omega}_{A}$, the spin connection
form ${\omega}_{A}^{S}$ can be obtained as
\begin{align}
{\omega}_{A}^{S} = \frac {1} {4} {{{\omega}_{A}}^{b}}_{c} {\not \!e}_{b}{\not
\!e}^{c}
=\frac {1} {2}{\not \!e}_{0}{\not \!e}^{2} \sinh{\theta} d\phi = \frac {1} {2}
{\not \!e}_{0}{\not \!e}^{2}\sinh{\theta}
\frac {{e}^{2}} {R\cosh{\theta}}
\end{align}

The intrinsic Dirac operator is then
\begin{align}
\not\!\! D &= {\not \!e}^{a}{D}_{a}= {\not \!e}^{0}\frac
{1}{R}\frac {\partial}{\partial \theta} +{\not \!e}^{2}\frac
{1}{R\cosh{\theta}}\frac {\partial}{\partial \phi} +\frac {1}
{2R}{\not \!e}^{0} \tanh{\theta}
\end{align}

\subsubsection{Extrinsic description}

\begin{definition}\label{extrinsicconnectiondef}
Let $M$ be given as a submanifold  in ${\mathbb{R}}^{n+1}$ of codimension 1,
equipped with the metric induced from a global
(translation invariant) metric on ${\mathbb{R}}^{n+1}$.
The {\bf Levi-Civita covariant spin derivative} (in an \underline{extrinsic}
basis) along a line parametrized by $\lambda$ is given by
\begin{align}
{\nabla}_{\frac {d} {d\lambda}}\psi = \frac {d\psi} {d\lambda} + \frac {1} {2}
\not\!\! n \frac {d\not\!\! n} {d\lambda}
\end{align}
\end{definition}

This may be used to define an extrinsic Dirac operator, as discussed in the next
subsection.

\subsubsection{Comparison of extrinsic and intrinsic calculations}

The intrinsic calculations proceed from the extrinsic point of view in a moving
frame ${e}_{\bullet}$, related to the
global frame ${b}_{\bullet}$ by orthogonal transformations. It is thus useful to
know the spin
matrices implementing those orthonormal transformations on spinor space.

The generators of rotations and boosts in spinor space can be constructed out of
the Clifford generators ${\gamma}_{i}$
by
\begin{align}
{\omega}_{ij} &= \frac {1} {4} [{\gamma}_{i}, {\gamma}_{j}]
\end{align}
The Clifford generators ${\gamma}_{i}$ satisfy
\begin{align}
[{\gamma}_{0},{\gamma}_{1}]&=2{\gamma}_{2},\\
[{\gamma}_{0},{\gamma}_{2}]&=-2{\gamma}_{1},\\
[{\gamma}_{1},{\gamma}_{2}]&=-2{\gamma}_{0}.
\end{align}

It follows for the generators of rotations and boosts on the spinor space

\begin{align}
[{\omega}_{01},{\omega}_{21}]&=-{\omega}_{02}\label{omega1},\\
[{\omega}_{02},{\omega}_{21}]&={\omega}_{01},\\
[{\omega}_{01},{\omega}_{02}]&=-{\omega}_{21}.\label{omega3}
\end{align}

These are the commutation relations of the Lie algebra ${sl}_{2}(\mathbb{R})$
(see Section \ref{SL2Rreptheory} for additional facts on
${sl}_{2}(\mathbb{R})$ and on the corresponding Lie group
${Sl}_{2}(\mathbb{R})$), the generators of (the double cover of)
the group of Lorentz transformations.

The boost and rotation spin matrices are:
\begin{align}
{S}_{01-boost}&=
\begin{pmatrix}
e^{\frac{\theta }{2}} & 0 \\
0 & e^{\frac{-\theta }{2}}
\end{pmatrix} &
{S}_{01-boost}^{-1}&=
\begin{pmatrix}
e^{\frac{-\theta }{2}} & 0 \\
0 & e^{\frac{\theta }{2}}
\end{pmatrix}\\
{S}_{02-boost}&=
\begin{pmatrix}
\cosh (\frac{\theta }{2}) &
\sinh (\frac{\theta }{2}) \\
\sinh (\frac{\theta }{2}) &
\cosh (\frac{\theta }{2})
\end{pmatrix} &
{S}_{02-boost}^{-1}&=
\begin{pmatrix}
\cosh (\frac{\theta }{2}) &
-\sinh (\frac{\theta }{2}) \\
-\sinh (\frac{\theta }{2}) &
\cosh (\frac{\theta }{2})
\end{pmatrix}\\
{S}_{12-rotation} &=
\begin{pmatrix}
\cos (\frac{\phi }{2}) &
\sin (\frac{\phi }{2}) \\
-\sin (\frac{\phi }{2}) &
\cos (\frac{\phi }{2})
\end{pmatrix} &
{S}_{12-rotation}^{-1} &=
\begin{pmatrix}
\cos (\frac{\phi }{2}) &
-\sin (\frac{\phi }{2}) \\
\sin (\frac{\phi }{2}) &
\cos (\frac{\phi }{2})
\end{pmatrix}
\end{align}

As the fact that the frames ${b}_{\bullet}$, ${e}_{\bullet}$ agree in a point
(see (\ref{agreeinpoint})) and from the knowledge of
the spin matrices,  the the transformation from ${b}_{\bullet}$ the global frame
of Min\-kow\-ski space to a local frame ${e}_{\bullet}$
can be calculated as ${S}_{\theta, \phi}={S}_{12-rotation}{S}_{01-boost}$

\begin{align}
{S}_{\theta ,\phi}&=
\begin{pmatrix}
e^{\frac{\theta }{2}}\,\cos (\frac{\phi }{2}) &
\frac{\sin (\frac{\phi }{2})}{e^{\frac{\theta }{2}}} \\
-\left( e^{\frac{\theta }{2}}\,\sin (\frac{\phi }{2}) \right)  &
\frac{\cos (\frac{\phi }{2})}{e^{\frac{\theta }{2}}}
\end{pmatrix} &
{S}_{\theta ,\phi}^{-1}&=
\begin{pmatrix}
\frac{\cos (\frac{\phi }{2})}{e^{\frac{\theta }{2}}} &
-\left( \frac{\sin (\frac{\phi }{2})}{e^{\frac{\theta }{2}}}\right)  \\
e^{\frac{\theta }{2}}\,\sin (\frac{\phi }{2}) &
e^{\frac{\theta }{2}}\,\cos (\frac{\phi }{2})
\end{pmatrix}
\end{align}

This is the spin matrix that allows an explicit comparison of  extrinsic and
intrinsic calculations.

At times, it is also useful to know the spin matrix implementing
the reflection exchanging  ${b}_{1}$ and ${b}_{2}$:

\begin{align}
{S}_{1\leftrightarrow 2}&={S}_{1\leftrightarrow 2}^{-1}=
\begin{pmatrix}
-1 & 1 \\
 1 & 1
\end{pmatrix}&
\end{align}

The full action of the symmetry group ${Sl}_{2}(\mathbb{R})$ on spinor fields is
given by the simultaneous action on the
spinor space and on the spacetime manifold. The full symmetry generators are
\begin{align}
{T}_{ij}&=-{L}_{ij}+{\omega}_{ij}
\end{align}
and satisfy the commutation relations

\begin{align}
[{T}_{01},{T}_{21}]&=-{T}_{02}\\
[{T}_{02},{T}_{21}]&={T}_{01}\\
[{T}_{01},{T}_{02}]&=-{T}_{21}
\end{align}

analogous to  (\ref{L1})-(\ref{L3}) and (\ref{omega1})-(\ref{omega3}). From the
point of view of representation theory, it is useful to
work with the rising and lowering operators
\begin{align}
{T}_{\pm} &= {T}_{01}\mp i{T}_{02},
\end{align}

satisfying

\begin{align}
[{T}_{\pm},{T}_{21}] &= -{T}_{02}\mp i{T}_{01}= \mp {T}_{\pm},\\
[{T}_{+},{T}_{-}] &= -2 i{T}_{21}.
\end{align}


\subsection{The Dirac operator on de Sitter space}

In the following, it is shown that the Dirac operator on
Minkowski space commutes with the generators of the Lorentz group
action on Minkowski space spinors. Then the Dirac operator on a
surface of codimension $1$ in flat space is defined and
reexpressed in terms of the intrinsic curvature of the surface
and calculated for spacelike hyperboloids in Minkowski space. It
is further shown that the Dirac operators on the hyperboloids
also commute with the generators of the Lorentz group.

\begin{lemma}
The Dirac operator $\not\!\!{D}_{M} = {\gamma}^{i}{\partial}_{i}$
commutes with the symmetry generators
${T}_{ij}=-{L}_{ij}+{\omega}_{ij}$.
\end{lemma}

\begin{proof}
By definition
\begin{align}
  {L}_{ij} &= {x}^{k} {g}_{k[i}{\partial}_{j]} &
  \text{with
  ${g}_{k[i}{\partial}_{j]}={g}_{ki}{\partial}_{j}-{g}_{kj}{\partial}_{i} $,} \\
  {\omega}_{ij} &= \frac {1} {4} [{\gamma}_{i},{\gamma}_{j}].&
\end{align}
A direct calculation gives then
\begin{align}\label{comsymDirac}
  [\not\!\!{D}_{M}, {T}_{ij}] &= -
  {\gamma}^{l}({\partial}_{l}{L}^{k}_{ij}){\partial}_{k}
  - [{\omega}_{ij},{\gamma}^{k}]{\partial}_{k} =
  \underbrace{\left( {\gamma}_{i}{\delta}_{j}^{k}-{\gamma}_{j}{\delta}_{i}^{k} -
  \frac {1} {4}
  [[{\gamma}_{i},{\gamma}_{j}],{\gamma}^{k}]\right)}_{\text{vanishes}}{\partial}_{k}
\end{align}
The coordinates of the commutator vanish for the following
reasons:

\begin{itemize}
  \item If $i=j$ then it vanishes due to antisymmetry in $i$, $j$;
  \item If $i$, $j$, $k$ are all different then the Kronecker
  $\delta$-s vanish and $[{\gamma}_{i},{\gamma}_{j}]$ commutes
  with ${\gamma}_{k}$, thus everything vanishes;
  \item If $k=i\neq j$ then ${\delta}_{j}^{k}$ vanishes and
\begin{align*}
  [[{\gamma}_{i}, {\gamma}_{j}],{\gamma}^{k}] &=
  2[{\gamma}_{i}{\gamma}_{j},{\gamma}^{k}]= -2 \{ {\gamma}_{i},
  {\gamma}^{k}\} {\gamma}_{j} = -4 {\delta}_{i}^{k}{\gamma}_{j},
\end{align*}
and the resulting term cancels with
$-{\gamma}_{j}{\delta}_{i}^{k}$;
  \item If $k=j\neq i $ then use antisymmetry in $i,j$ to
  transform it into the previous case.
\end{itemize}
\end{proof}

\begin{definition}
(The Dirac operator on a surface of codimension 1.) Let ${e}_{A}$
be a vielbein on the $n-1$-dimensional surface in $n$-dimensional
flat space and let ${\gamma}^{\bullet}$ be global generators of
the global Clifford algebra. Then the Dirac operator on the
surface is given by
\begin{align}\label{Diracextrinsicdef}
\not\!\!{D}&= {\gamma}^{A}{e}_{A}^{B}{\nabla}_{B}
\end{align}
where ${\nabla}_{\bullet}$ is the extrinsic connection of
Definition \ref{extrinsicconnectiondef}
\end{definition}

A direct calculation, using the definition
${{K}_{A}}^{B}:={e}_{A}^{C}{\partial}_{C}{n}^{B}$ of the
extrinsic curvature ${{K}_{A}}^{B}$ gives
\begin{align}\label{Diracextrinsiccalc}
  \not\!\!{D}\psi&= {\gamma}^{A}{e}_{A}^{B}{\partial}_{B}\psi +
  \frac {1} {2}{\gamma}^{A}{e}^{B}_{A}\not\!{n}{\partial}_{B}\not\!{n}\psi = \\
  &= \not\!{e}^{B}{\partial}_{B}\psi + \frac {1} {2}
  \underbrace{\not\!{e}^{B}\not\!{n}}_{\text{anticommute}}{\partial}_{B}\not\!{n}\psi
  = \\
  &= \not\!{e}^{B}{\partial}_{B}\psi - \frac {1}
  {2}\not\!{n}\underbrace{\not\!{e}^{B}{\partial}_{B}\not\!{n}}_{\not\!{e}^{A}{K}_{AB}\not\!{e}^{B}}=\\
  &= \not\!{e}^{B}{\partial}_{B}\psi - \frac {1}
  {2}{{K}_{A}}^{A}\not\!{n}
\end{align}

In the following lemma, the needed extrinsic curvature is calculated:

\begin{lemma}\label{extrinsiccurvaturecalculation}
The extrinsic curvature tensor ${{K}_{A}}^{B}$ and its trace
${{K}_{A}}^{A}$ of the hyperboloid $-{({x}^{0})}^{2} +
\sum_{i}{({x}^{i})}^{2}={R}^{2}$ in $n$-dimensional Minkowski space are
given by
\begin{align}\label{extrinsiccurvature}
  {{K}_{A}}^{B} &= \frac {1} {R}{\delta}_{A}^{B}  \\
  {{K}_{A}}^{A} &= \frac {n-1} {R}
\end{align}
\end{lemma}

\begin{proof}
Due to the transitive action of the symmetries on the
hyperboloid, it is sufficient to calculate the formulae in one
point only which may conveniently be chosen to be
\begin{align*}
  {x}^{\alpha} = (0, R, 0, 0, \ldots , 0) \\
  \intertext{with normalized coordinates around this point:}
  {x}^{A} = \frac {1} {R}({x}^{0}, {x}^{2}, {x}^{3}, \ldots ,
  {x}^{n-1})\\
  \intertext{and with the normal vector:}
  {n}^{\alpha} = \frac {1} {R}({x}^{0}, {x}^{1} {x}^{2}, {x}^{3}, \ldots ,
  {x}^{n-1})
\end{align*}
Then a simple calculation gives
\begin{align}
    {{K}_{A}}^{B} &= \frac {d {n}^{B}}{d{x}^{A}}=\frac {1} {R}{\delta}_{A}^{B}
\end{align}
\end{proof}

\begin{corollary}
On the hyperboloid $-{({x}^{0})}^{2} + \sum_{i}{({x}^{i})}^{2}={R}^{2}$
in $n$-dimensional Min\-kow\-ski space, the Dirac operator
$\not\!\!{D}$ is given by
\begin{align}\label{deSitterDirac}
  \not\!\!{D} &= {\not\! e}^{B}{\partial}_{B}-\frac {n-1} {2R}
  \not\!{n}
\end{align}
\end{corollary}

\begin{corollary}
On the hyperboloid $-{({x}^{0})}^{2} + {({x}^{i})}^{2}={R}^{2}$
in $n$-dimensional Min\-kow\-ski space, the Dirac operator
$\not\!\!{D}$ can be induced from the Dirac operator
${\not\!\!{D}}_{M}$ on Min\-kow\-ski space by extending the spinors
outside the hyperboloid by $(\text{value on hyperboloid})\times
{e}^{-\frac{n-1}{2R}(r-R)}$ along the rays $\lambda {x}^{\alpha}$,
with $r$ being the spacetime interval as measured form the origin.
\end{corollary}

\begin{corollary}
The Dirac operator $\not\!\!{D}$ on the hyperboloid commutes with
the symmetry generators ${T}_{ij}$.
\end{corollary}

\begin{proof}
Note first that the symmetry generators ${T}_{ij}$ preserve the
hyperboloid \\
$-{({x}^{0})}^{2}+{({x}^{i})}^{2}={R}^{2}$ in
$n$-dimensional Minkowski space. The Minkowski space Dirac
operator ${\not\!\!{D}}_{M}$ commutes with the symmetry
generators ${T}_{ij}$. From the previous corollary,
${\not\!\!{D}}$ can be understood to be a restriction of
${\not\!\!{D}}_{M}$ to a subset of functions behaving radially as
${e}^{-\frac{n-1}{2R}(r-R)}$.
\end{proof}



\subsection{The implementation of symmetries on the spinor field on de Sitter}

The de Sitter hyperboloid embedded in Minkowski space is left invariant under
the action of the Lorentz group and thus an
 an action of the Lorentz group can be given on spinor fields
on this hyperboloid.

A representation of the Lorentz group can be given by restricting
to the spinor fields $\psi$ which are solutions of the Dirac
equation:
\begin{align}\label{Diraceq}
  (\not\!\!{D}-m)\psi =0
\end{align}
This is reasonable, since the Dirac operator commutes with the
generators of the Lorentz group and since the space of solutions
of the Dirac equation is equipped with an invariant sesquilinear
positive definite inner product:
\begin{align}\label{phasespaceproduct}
  ({\psi}_{1}, {\psi}_{2}) &=
  \int_{\Sigma}{\bar{{\psi}_{1}}{\gamma}^{A}{\psi}_{2}{d}_{A}\Sigma},
\end{align}
with $\Sigma $ being an arbitrary spacelike Cauchy surface on the
hyperboloid.

The Hilbert space of the representation is then the classical
phase space of spinor fields, the space of initial values of the
spinor field on a Cauchy surface satisfying the Dirac equation. In
order to give the action of the Lorentz group on the initial
values, the Dirac equation has to be used.

The Cauchy hypersurface will be taken a sphere ${S}^{n-2}$ on the
hyperboloid at ${x}^{0}= constant$. It has a normal vector
$\vec{n}=\frac {1} {R} ({x}^{0}, {x}^{1}, \ldots , {x}^{n-1})$
and an (unnormalized) spacial normal vector $\vec{r}= (0, {x}^{1},
\ldots , {x}^{n-1})$. The time evolution vector will be taken as
\begin{align}
\frac {\partial} {\partial t} := {e}_{0} = \frac {1} {R} \frac
{\partial} {\partial \theta} &= \cosh{\theta} \frac {\partial}
{\partial {x}^{0}} + \sinh{\theta} \frac {\vec{r}} {\mid
\vec{r}\mid} =\\
&= \frac {1} {\cosh{\theta}} \left( \frac {\partial} {\partial {x}^{0}} +
\sinh{\theta} \vec{n} \right).
\end{align}

The Dirac operator can then be written as
\begin{align}
\not\!\!{D} &= {\not\!{e}}^{0} {\partial}_{t} + {\not\!{e}}^{I} {\partial}_{I} -
\frac {n-1} {2R} {\not\!{n}},\\
\intertext{and using the Dirac equation}
(\not\!\!{D} -m)\psi &=0,\\
\intertext{the infinitesimal time evolution can be given:}
{\partial}_{t} &= [{\not\!{e}}_{J}, {\not\!{e}}_{0}] {g}^{JI} {\partial}_{I} +
\frac {n-1} {2R} {\not\!{e}}_{0}{\not\!{n}} + m{\not\!{e}}_{0} =\\
&= \underbrace{\frac {\not\!{n}} {R\cosh{\theta}}{L}_{21} +\frac {1} {R}
{\not\!{e}}_{0}{\not\!{n}} + m{\not\!{e}}_{0}}_{\text{for 2-dim. de Sitter
space}}
\end{align}

The generators ${T}_{ij}$ of the Lorentz transformations in 3-dimensional
Minkowski space acting on spinor fields on the de Sitter space of radius $R$ can
now be expressed
in terms of ${\partial}_{t}$, ${\partial}_{\phi}$ which in turn can be
expressed
in terms of the Cartan subalgebra generator $T_{21}$, eliminating
all derivatives in the formulae.

\begin{align}
\frac {\partial} {\partial t} &= \frac {\not\!{n}} {R\cosh{\theta}} \left(
{L}_{21}-{\not\!{e}}_{0}\cosh{\theta} \right) + m {\not\!{e}}_{0}=\\
&=\frac {\not\!{n}} {R\cosh{\theta}} \left(-{T}_{21}+\frac {1}
{2}{\gamma}_{0}-{\not\!{e}}_{0}\cosh{\theta} \right) + m
{\not\!{e}}_{0}\label{timederiv}\\
\frac {\partial} {\partial \phi} &= -{T}_{21}+\frac {1}
{2}{\gamma}_{0}
\end{align}

\begin{align}
{\omega}_{21}&=\frac {1} {2}{\gamma}_{0}\\
{\omega}_{02}&=-\frac {1} {2}{\gamma}_{1}\\
{\omega}_{01}&=\frac {1} {2}{\gamma}_{2}
\end{align}

\begin{align}
{L}_{21}&=\frac {\partial} {\partial \phi}\\
{L}_{02}&= R \sin{\phi}\frac {\partial} {\partial t} + \cos{\phi}\tanh{\theta}
\frac {\partial} {\partial \phi}\\
{L}_{01}&= \cos{\phi}\frac {\partial} {\partial \theta} -
\sin{\phi}\tanh{\theta} \frac {\partial} {\partial \phi}
\end{align}

\begin{align}
{T}_{21}&=-\frac {\partial} {\partial \phi}+\frac {1} {2}{\gamma}_{0}\\
{T}_{02}&= -R \sin{\phi}\frac {\partial} {\partial t} -
\cos{\phi}\tanh{\theta} \frac {\partial} {\partial \phi}-\frac {1}
{2}{\gamma}_{1}\\
{T}_{01}&= -R \cos{\phi}\frac {\partial} {\partial t} +
\sin{\phi}\tanh{\theta} \frac {\partial} {\partial \phi}+ \frac
{1} {2}{\gamma}_{2}
\end{align}

There are now two natural bases in the space of spinors on
${S}^{n-2}({x}^{0})={S}^{1}({x}^{0})$, namely the eigenbasis $\mid L:
n,\pm\rangle$
of ${L}_{21}$ and the eigenbasis $\mid T: n,\pm\rangle$ of
${T}_{21}$. The first choice looks to be simpler, if written in
the local basis of eigenvectors of ${\gamma}_{0}$:

\begin{align}\label{L-eigenvectors}
 \mid L: n,+\rangle &= \begin{pmatrix}
 {e}^{in\phi} \\
 0
\end{pmatrix} &
\mid L: n,-\rangle &=\begin{pmatrix}
0 \\
{e}^{in\phi}
\end{pmatrix}\\
\end{align}
with $n$ being integers $\ldots, -2,-1,0,+1,+2,\ldots$.

and is maybe somewhat easier to visualize. Compare it with the
second basis:
\begin{align}\label{T-eigenvectors}
 \mid T: n,+\rangle &= \begin{pmatrix}
 {e}^{-i(n-\frac {1} {2})\phi} \\
 0
\end{pmatrix} &
\mid T: n,-\rangle &=\begin{pmatrix}
0 \\
{e}^{-i(n+\frac {1} {2})\phi}
\end{pmatrix}
\end{align}
with $n$ being half-integers $\ldots, -\frac {5} {2}, -\frac {3}
{2},-\frac {1} {2},+\frac {1} {2},+\frac {3} {2},+\frac {5}
{2},\ldots$.

However it is more appropriate to use the second one since
\begin{itemize}
  \item it is the full symmetry generators ${T}_{ij}$ that are to
  play the key role in the representation theory of the symmetry
  group,
  \item the generators ${L}_{ij}$ may not exist on the spinor
  bundle of a higher dimensional example, due to its
  non-triviality and is thus not the right concept to be used. To
  give a meaning to ${L}_{ij}$ one would have to enlarge the
  spinor bundle, e.g. by adding to it an inverse bundle.
\end{itemize}

\begin{note}
For the reasons given above, the calculations of the action of the symmetry
generators ${T}_{\bullet}$ will be continued in the  basis $\mid T:
n,\pm\rangle$. But for
completeness, analogous results in the basis $\mid L: n,\pm\rangle$ are stated
here, in advance:

\begin{align*}
{T}_{21}\mid L: n,+\rangle &=\frac {i} {2} (1-2n) \mid L: n,+\rangle \\
{T}_{21}\mid L: n,-\rangle &=\frac {i} {2} (-1-2n) \mid L: n,-\rangle \\
{T}_{01}\mid L: n,+\rangle &=\frac {1} {2}
\left[
  n     \mid L: n,-\rangle
+ (n-1) \mid L: n-2,-\rangle
- is    \mid L: n+1,+\rangle
- is    \mid L: n-1,+\rangle
\right]\\
{T}_{01}\mid L: n,-\rangle &=\frac {1} {2}
\left[
-  n    \mid L: n,+\rangle
- (n+1) \mid L: n+2,+\rangle
+ is    \mid L: n+1,-\rangle
+ is    \mid L: n-1,-\rangle
\right]\\
{T}_{02}\mid L: n,+\rangle &=\frac {i} {2}
\left[
- n     \mid L: n,-\rangle
+ (n-1) \mid L: n-2,-\rangle
+ is    \mid L: n+1,+\rangle
- is    \mid L: n-1,+\rangle
\right]\\
{T}_{02}\mid L: n,-\rangle &=\frac {i} {2}
\left[
-  n    \mid L: n,+\rangle
+ (n+1) \mid L: n+2,+\rangle
- is    \mid L: n+1,-\rangle
+ is    \mid L: n-1,-\rangle
\right]\\
\end{align*}

\begin{align}
{T}_{+}\mid L: n,+\rangle &= (n-1) \mid L: n-2,-\rangle - is \mid L:
n-1,+\rangle\\
{T}_{+}\mid L: n,-\rangle &= -n \mid L: n,+\rangle + is \mid L: n-1,-\rangle\\
{T}_{-}\mid L: n,+\rangle &= n \mid L: n,-\rangle - is \mid L: n+1,+\rangle\\
{T}_{-}\mid L: n,-\rangle &= -(n+1) \mid L: n+2,+\rangle + is
\mid L: n+1,-\rangle
\end{align}

\end{note}

We now proceed with the calculation of the action of the
${sl}_{2}(\mathbb{R})$-symmetries (Lorentz symmetries) on the space
 of solutions of the Dirac equation in the basis $\mid T: n, \bullet\rangle$.

By definition,
\begin{align}\label{T12action}
   {T}_{21}\mid T: n,\pm\rangle = in \mid T: n, \pm\rangle.
\end{align}

\begin{align}
  {\not\!{e}}_{0} &= \cosh{\theta}\underbrace{{\not\!{b}}^{0}}_{={\gamma}^{0}}+
  \sinh{\theta} \frac {\vec{{\not\!{r}}}} {R}\\
  {\not\!{n}} &= \sinh{\theta}{\not\!{b}}^{0}+ \cosh{\theta} \frac
  {\vec{{\not\!{r}}}} {R}\\
  \frac {\vec{{\not\!{r}}}} {R} &:= \frac {{x}^{1}}
  {R}{\gamma}_{1}+\frac {{x}^{2}} {R}{\gamma}_{2}=
  {e}^{i\phi}({\gamma}_{1}-i{\gamma}_{2})+{e}^{-i\phi}({\gamma}_{1}+i{\gamma}_{2})=\\
  &=-i{e}^{i\phi}
  \begin{pmatrix}
   0&1 \\
   0&0
  \end{pmatrix}
  +i{e}^{-i\phi}
  \begin{pmatrix}
   0&0 \\
   1&0
  \end{pmatrix}
\end{align}
The actions on the basis are
\begin{align}
\frac {\vec{{\not\!{r}}}} {R} \mid T: n,\pm \rangle &:= \pm i\mid
T: n,\mp \rangle\label{a1}\\
{\not\!{e}}_{0} \mid T: n,\pm \rangle &:= \pm i\cosh{\theta} \mid
T: n,\pm \rangle
\pm i\sinh{\theta} \mid T: n,\mp \rangle\label{a2}\\
{\not\!{n}} \mid T: n,\pm \rangle &:= \pm i\sinh{\theta} \mid T: n,\pm
\rangle \pm i\cosh{\theta} \mid n,\mp \rangle\label{a3}
\end{align}

A direct calculation of $\frac {\partial} {\partial\theta}=R\frac {\partial}
{\partial t}$ from (\ref{timederiv}) using
(\ref{a1})-(\ref{a3}) gives
\begin{align}
\frac {\partial} {\partial\theta} \mid T: n, \pm\rangle &= \left( (\pm n -\frac
{1} {2})\tanh{\theta} \pm iRm
\cosh{\theta}\right) \mid T: n, \pm\rangle\\
&+ \left( (\pm n +\frac {1} {2}) \pm iRm
\sinh{\theta}\right) \mid T: n, \mp\rangle
\end{align}

\begin{align}
{T}_{+} \mid
T: n,\pm\rangle&= \left[ (n+\frac {1} {2})\tanh{\theta}(1\mp 1) \mp
iRm\cosh{\theta}\right] \mid
T: n+1,\pm\rangle\\
&\mp \left[ n+\frac {1} {2} + iRm\sinh{\theta} \right] \mid
T: n+1,\mp\rangle\\
{T}_{-} \mid T: n,\pm\rangle&= \left[ -(n-\frac {1} {2})\tanh{\theta}(1\pm 1)
\mp iRm\cosh{\theta}\right] \mid
T: n-1,\pm\rangle\\
&\mp \left[ n-\frac {1} {2} + iRm\sinh{\theta} \right] \mid T:
n-1,\mp\rangle
\end{align}

In matrix notation, on the subspace $\mid T: n,\bullet\rangle$, one gets

\begin{align}
{T}_{+}(n) &=
\begin{pmatrix}
   -iRm\cosh{\theta}& n+\frac {1} {2} + iRm\sinh{\theta}\\
   -(n+\frac {1} {2} + iRm\sinh{\theta})&2(n+\frac {1} {2})\tanh{\theta} +
   iRm\cosh{\theta}
\end{pmatrix}u\\
{T}_{-}(n) &=
\begin{pmatrix}
   -2(n-\frac {1} {2})\tanh{\theta} -iRm\cosh{\theta}& n-\frac {1} {2} +
   iRm\sinh{\theta}\\
   -(n-\frac {1} {2} + iRm\sinh{\theta})&iRm\cosh{\theta}
\end{pmatrix}{u}^{\ast}
\intertext{and the time vector ${\not\!{e}}_{0}$ is given by}
{\not\!{e}}_{0} &= i
\begin{pmatrix}
   \cosh{\theta}& -\sinh{\theta}\\
   \sinh{\theta}& -\cosh{\theta}
\end{pmatrix}
\end{align}

These matrices are understood to be given with respect to the
basis $\mid T: n, \pm\rangle$. This basis is however not
orthonormal. In order to find an orthogonal basis which would be
suitable for comparison with representation-theoretic
calculations, the structure of the inner product on the space of
spinor fields has to be clarified.

Spacetime spinor fields are equipped with a hermitean inner
product, the Dirac product $B(\bullet,\bullet)$ which is given as
the intertwiner $B$
\begin{align}
  {{\gamma}_{\mu}}^{+} B  &= - B  {\gamma}_{\mu} \\
  B &= \begin{pmatrix}  -i&0 \\ 0 & i \end{pmatrix}\\
  \intertext{so that the inner product can be calculated as}
  B(\psi,\varphi)&= \begin{pmatrix}  \bar{{\psi}_{1}}&
  \bar{{\psi}_{2}}\end{pmatrix}
  \begin{pmatrix}  -i&0 \\ 0 & i \end{pmatrix}
  \begin{pmatrix}   {{\varphi}_{1}} \\  {{\varphi}_{2}} \end{pmatrix}
\end{align}

\begin{note}
There is a second possibility to choose a hermitean inner product
$A$ by taking the opposite sign in the intertwining relation:
\begin{align}
   {{\gamma}_{\mu}}^{+} A  &= + A  {\gamma}_{\mu}
\end{align}
But this is irreconcilable with other choices already made or to be required, in
particular
\begin{itemize}
\item The Dirac operator was defined to be $\dirac =
{\gamma}^{\mu}{\partial}_{\mu}$ rather than
$i{\gamma}^{\mu}{\partial}_{\mu}$.
\item The Dirac equation was chosen to be $(\dirac -m)\psi =0$.
\item The Dirac operator is required to be formally selfadjoint,
\begin{align*}
  B(\psi,\dirac\varphi)-B(\dirac\psi,\varphi)={\partial}_{\mu}
  B(\psi,{\gamma}^{\mu}\varphi).
\end{align*}
\item The physical time directions given by the metric are negative definite
(rather than positive definite), i.e.,
${{p}_{\mu}p^{\mu}}=-{m}^{2}$ for an approximate plane wave (rather than
${{p}_{\mu}p^{\mu}}=+{m}^{2}$)
\end{itemize}
\end{note}

It follows from the global hyperbolicity of the spacetime and the
formal selfadjointness of the Dirac operator that the following is
a well defined hermitean inner product on the space of solutions of the Dirac
equation,
independent of the spacelike Cauchy surface $\Sigma$, see
(\ref{phasespaceproduct}):
\begin{align}
  (\psi,\varphi)=\int_{{S}^{1}(\theta =\text{constant)}}{B(\psi (\phi)
  ,{e}^{0}\varphi (\phi))}{d\phi}
\end{align}

With respect to this inner product,
the normalized eigenvectors ${e}_{\pm i}$ of  ${\not\!{e}}_{0}$ for the
eigenvalues $\pm i$ are
\begin{align}
{e}_{\pm i}&= \frac {1} {\sqrt{2\cosh{\theta}\pm2}}
\begin{pmatrix}
\cosh{\theta}\pm 1\\
\sinh{\theta}
\end{pmatrix},
\end{align}

and in this {\bf orthogonal eigenbasis of} ${\not\!{e}}_{0}$ we have
\begin{align}
{\not\!{e}}_{0}&=
  \begin{pmatrix}
  i & 0\\
  0 & -i
  \end{pmatrix},\label{sq1}\\
\intertext{ }
{\not\!{e}}_{2}&=
  \begin{pmatrix}
  0 & 1\\
  1 & 0
  \end{pmatrix},\\
\intertext{ }
{T}_{+}(n)&=
  \begin{pmatrix}
  -iRm + \left( n+\frac {1} {2}\right)\tanh{\theta}       & \frac {1}
  {\cosh{\theta}}\left( n+\frac {1} {2}\right)\\
  -\frac {1} {\cosh{\theta}}\left( n+\frac {1} {2}\right) & iRm + \left( n+\frac
  {1} {2}\right)\tanh{\theta}
  \end{pmatrix},\\
\intertext{ }
{T}_{-}(n)&=
  \begin{pmatrix}
  -iRm - \left( n-\frac {1} {2}\right)\tanh{\theta}       & \frac {1}
  {\cosh{\theta}}\left( n-\frac {1} {2}\right)\\
  -\frac {1} {\cosh{\theta}}\left( n-\frac {1} {2}\right) & iRm - \left( n-\frac
  {1} {2}\right)\tanh{\theta}
  \end{pmatrix},\\
  \intertext{and in particular}
{u}_{+}= {T}_{+}(+\frac {1} {2})&=
  \begin{pmatrix}
  -iRm + \tanh{\theta}       & \frac {1} {\cosh{\theta}}\\
  -\frac {1} {\cosh{\theta}} & iRm + \tanh{\theta}
  \end{pmatrix},\\
\intertext{ }
{u}_{-}= {T}_{-}(-\frac {1} {2})&=
  \begin{pmatrix}
  -iRm + \tanh{\theta}       & -\frac {1} {\cosh{\theta}}\\
   \frac {1} {\cosh{\theta}} & iRm + \tanh{\theta}
  \end{pmatrix}.\label{sq6}
\end{align}

These data may be compared with the data of the de Sitter spectral quadruples
obtained in Section \ref{deSitterrecover}


\end{document}